\documentclass[a4paper,fleqn,usenatbib]{mnras}

\usepackage{newtxtext,newtxmath}

\usepackage[T1]{fontenc}
\usepackage{ae,aecompl}

\usepackage{graphicx}	
\usepackage{amsmath}	
\usepackage{amssymb}	
\usepackage{algorithm}
\usepackage[noend]{algpseudocode}

\title[Using bootstrap to assess uncertainties of VLBI results]{Using bootstrap to assess uncertainties of VLBI results I. The method and image-based errors}

\author[I.N. Pashchenko]{
Ilya N. Pashchenko$^{1}$\thanks{E-mail: in4pashchenko@gmail.com (INP)}
\\
$^{1}$Astro Space Center of Lebedev Physical Institute, Profsoyuznaya Str.
84/32, 117997 Moscow, Russia\\
}

\date{Accepted 2018 September 24. Received 2018 September 19; in original form 2017 May 17}

\pubyear{2018}

\begin{document}
\label{firstpage}
\pagerange{\pageref{firstpage}--\pageref{lastpage}}
\maketitle

\begin{abstract}
Very Long Baseline Interferometric (VLBI) observations of quasar jets enable one to measure many theoretically expected effects. Estimating  the significance of observational findings is complicated by the correlated noise in the image plane. A reliable and well justified approach to estimate the uncertainties of VLBI results is needed as well as significance testing criteria. We propose to use bootstrap for both tasks. Using  simulations we find that bootstrap-based errors for the full intensity, rotation measure, and spectral index maps have coverage closer to the nominal values than conventionally obtained errors. The proposed method naturally takes into account heterogeneous interferometric arrays (such as Space VLBI) and can be easily extended to account for instrumental calibration factors.
\end{abstract}




\section{Introduction}
\label{sec:intro}



Very Long Baseline Interferometry (VLBI) is a powerful method for exploring the most compact emitting celestial objects.
It allows one to investigate the structure of relativistic quasar
jets \citep[e.g.][]{bllacs,nrao150}, their magnetic fields \citep[e.g.][]{mojave8rm,bllacrecollimation,bllacalfven,evk2017}, particle content \citep[e.g.][]{particles} etc. with the highest angular
resolution \citep{ra}. Typical VLBI results are images (distribution of Stokes parameters or their combinations, e.g. fractional linear polarization, spectral
index or Rotation Measure (RM) map) or direct models of the interferometric visibilities, which are
conventionally represented by several simple components \citep[e.g. circular or elliptical gaussians,][]{nonimagedataanalisis}.

Despite the great wealth of information obtained using VLBI, the uncertainties of the VLBI results, their robustness and statistical significance remain an open issue. This is because the procedure for VLBI data processing is very complex and includes numerous non-linear transformations. Reliable uncertainty estimates are crucial for testing the significance of the obtained results. At the same time even the most recent Space VLBI results with the highest angular resolution \citep[e.g.][]{2018NatAs...2..472G} were obtained using imaging algorithm that lacks uncertainty output. 

In this paper we propose to use bootstrap --- a well known method of assessing the uncertainties --- to estimate the errors in the VLBI images and their combinations.

In Section~\ref{sec:convmethods} we review conventional methods for estimating errors, their possible shortcomings, and those earlier studies that attempted to address these issues. In Section~\ref{sec:method} we describe the methodology, present the algorithm for the VLBI case, and outline the possibilities for extending. In Section~\ref{sec:simulations} we test the method using simulations with artificially generated data sets and compare our results with the conventional method for estimating image errors. In Section~\ref{sec:discussion} we discuss the limitations and advantages of using bootstrap for estimating uncertainties of VLBI results. We summarize our findings in Section~\ref{sec:conclusions}. We apply a real-life test to analyze the significance of RM gradients in Appendix~\ref{sec:real}.

\section{Conventional methods}
\label{sec:convmethods}

\subsection{Quantifying the image uncertainty}

An interferometer measures the Fourier components of the spatial spectra of the source (i.e. interferometric visibility function or visibilities) at certain points in the spatial frequency plane ($(u,v)$-plane). To obtain the image of the source one needs to deconvolve the Fourier Transform (FT) of the observed visibilities (\textit{dirty image}). The most widely used deconvolution algorithm is \texttt{CLEAN} \citep[][also its modern modifications, e.g Clark \texttt{CLEAN} \citealt{clarkclean}, multi-scale \texttt{CLEAN} \citealt{msclean}]{hogbom}. It effectively fits
the observed visibilities with the FT of a number of delta functions \citep{schwarz}. The output of the algorithm is the model of the observed visibility function consisting of a list of delta functions in the image plane (called \texttt{CLEAN} components, $\rm CC$). Its Fourier Transform equals the observed visibilities within errors. 

Due to using point sources for representing the source structure \texttt{CLEAN} does extrapolation in $(u,v)$-plane. \citep{briggs} found that this extrapolation is done noticeably incorrect. To exclude the information obtained from extrapolated high spatial frequencies the CCs are convolved with a Gaussian beam. The result of the convolution is called a \textit{CLEAN image}. The residual image left after deconvolution is traditionally added back to the \texttt{CLEAN} image to account for possible uncleaned flux and to estimate the noise \citep{briggs}.

   \begin{figure}
   \centering
   \includegraphics[width=\columnwidth, trim=0.3cm 0.5cm 0.3cm 0.3cm]{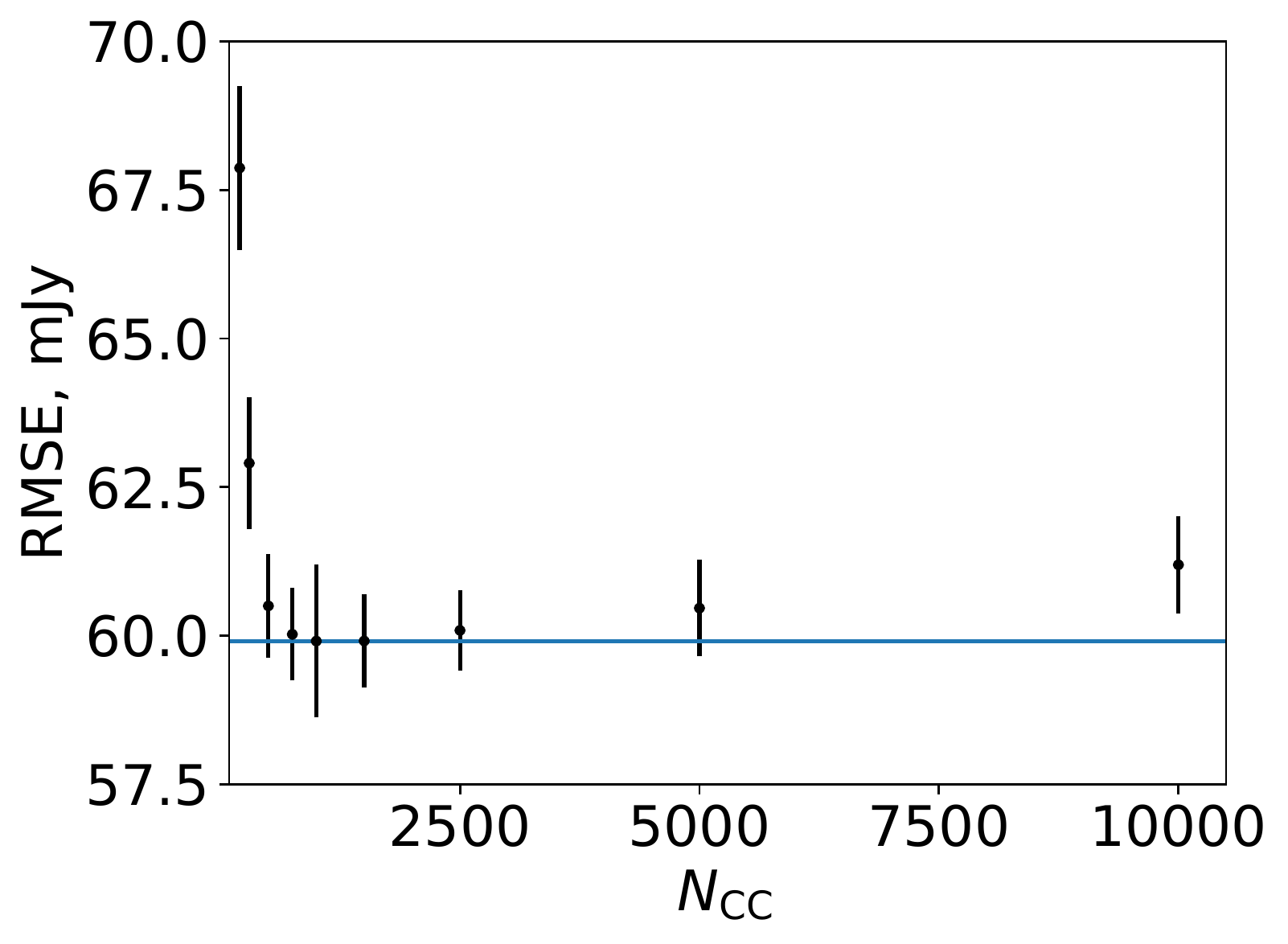}
      \caption{Cross-validation (CV) score vs. number of \texttt{CLEAN} iterations for source 0055$+$300.}
       \label{cv_score}
   \end{figure}

   \begin{figure}
   \centering
   \includegraphics[width=6cm, trim=0.3cm 0.3cm 0.3cm 0.3cm]{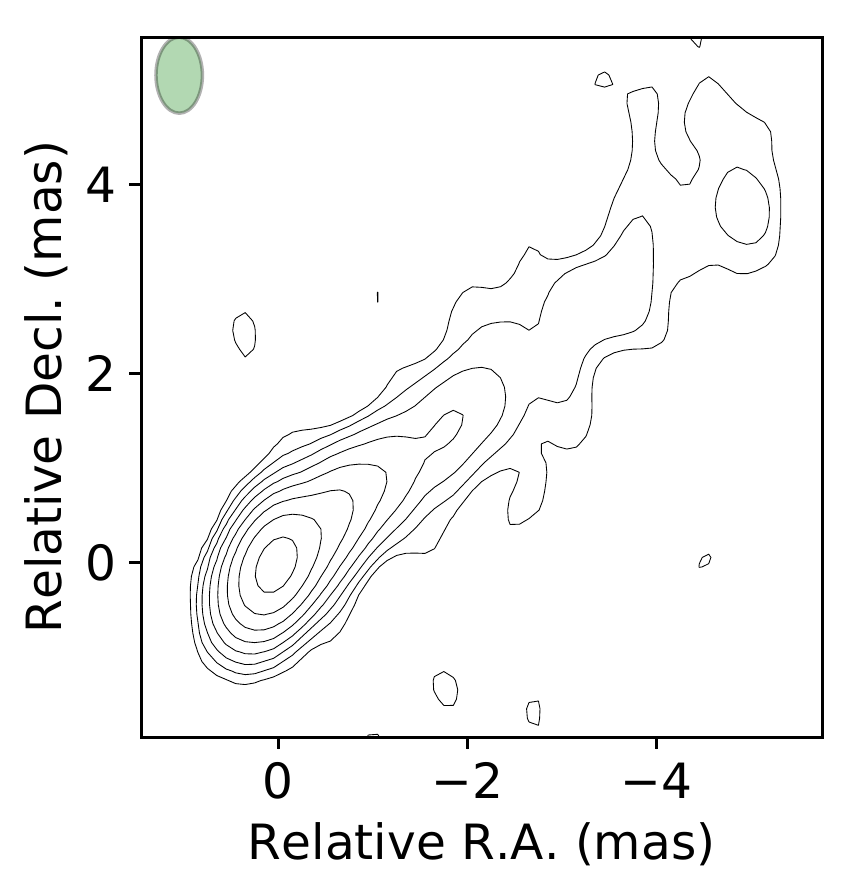}
      \caption{\texttt{CLEAN} image of source 0050$+$300 obtained using the number of iterations suggested by CV. The green ellipse in the image show the synthesized beam. Contours are plotted with a factor of 2. The lowest contour value is 1.9 mJy/beam.}
       \label{cv_map}
   \end{figure}
   
   \begin{figure}
   \centering
   \includegraphics[width=\columnwidth,trim=0.3cm 0.9cm 0.3cm 0.3cm]{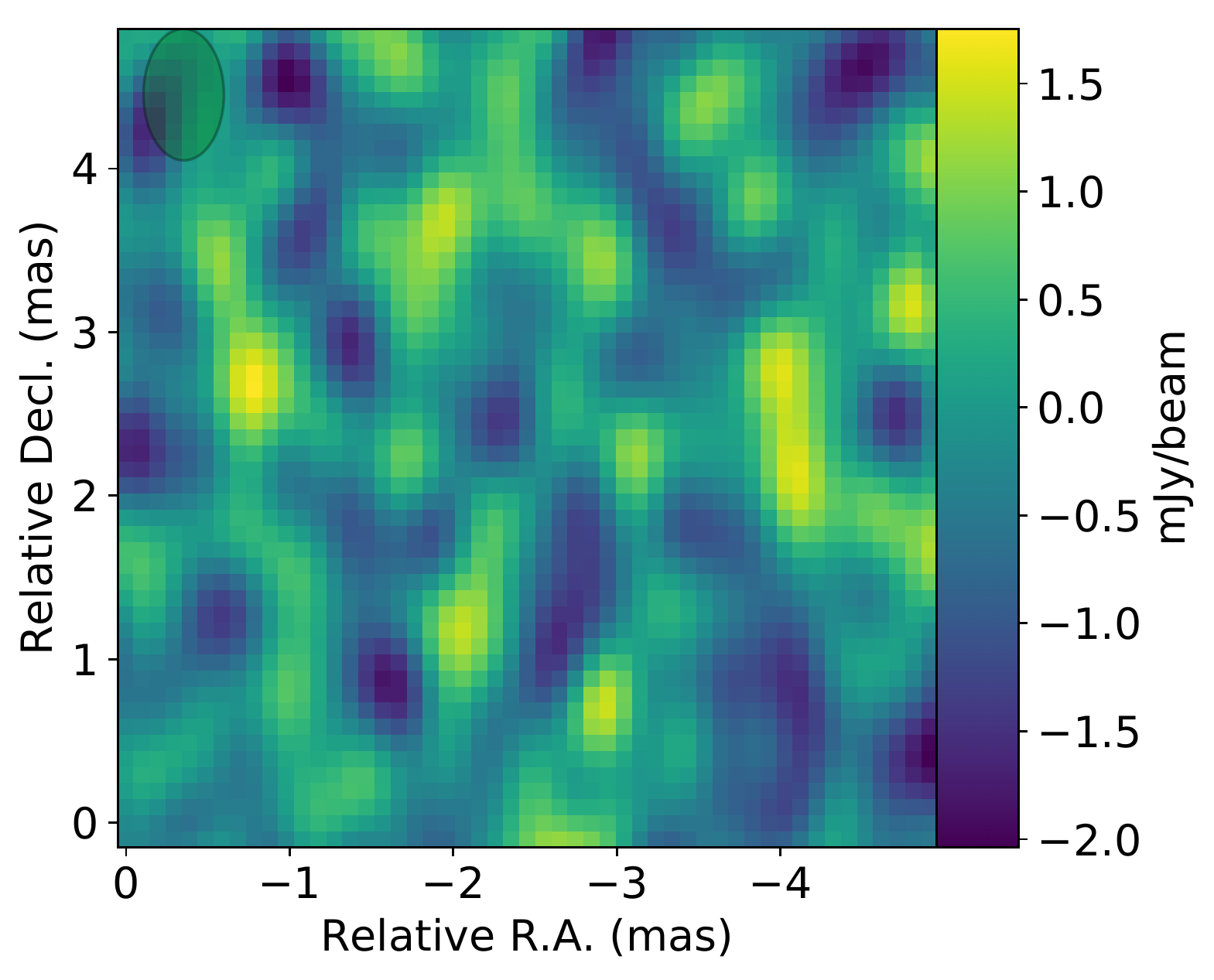}
      \caption{Part of the residuals image at the distance of 1 arcsecond from the CLEANing window. The green ellipse in the image show the synthesized beam.}
       \label{res}
   \end{figure}

It is conventional to describe the random (i.e. thermal noise) uncertainty of the flux values of the image pixels by the residual per-pixel $rms$ \citep{fomalont}. It is usually measured at some emission-free area of the region being \texttt{CLEAN}ed or even at some area distant from the \texttt{CLEAN}-region (e.g. 1 arcsec for VLBI maps). This assumes that the $rms$ of the residual image reflects the uncertainty of the CCs flux. At least two points should be considered at this step. First, one should choose the ``right'' \texttt{CLEAN} depth since \texttt{CLEAN}ing too deeply will result in fitting the noise. Then the $rms$-based errors will underestimate the errors of the CCs flux. Second, even with the ``best'' \texttt{CLEAN} depth it is nontrivial to
estimate the $rms$ of the residuals because they are correlated on the scale of a dirty beam\footnote{Dirty beam is the FT of the $(u,v)$-plane sampling function that equals one in the sampled points and zero elsewhere. A Dirty image is the convolution of the true image with the dirty beam. Poor coverage of $(u,v)$-plane results in a dirty beam with large sidelobes.}. This correlation is due to the convolution of noise with the dirty beam (when one calculates the residual $rms$ at an area distant from the \texttt{CLEAN}-region area) and the subtraction of the correlated values during \texttt{CLEAN} (when one calculates the residual $rms$ inside the \texttt{CLEAN}-region). We illustrate this by choosing the "best" \texttt{CLEAN} depth using the Cross-Validation\footnote{Cross-Validation estimates the prediction performance of the model on the independent (not used in fitting) data set.} (CV) procedure \citep[][Fig.~\ref{cv_score}]{esl}, which uses the root-mean-squared error (\textit{RMSE}) as a performance metric. We use the MOJAVE \citep{mojave6kinematicsanalysis} calibrated data for source 0050$+$300 at 15 GHz. The residuals left after \texttt{CLEAN} are correlated as can be clearly seen from Figure~\ref{res}. Thus to calculate the $rms$ one has to estimate the covariance matrix of the residuals. \cite{coughlangabuzda} attempt to address this problem by approximating the uncertainty of the individual pixel flux with the flux error of the \texttt{CLEAN} components falling within a given pixel and propagating the error of a sum of beam-convolved CCs. However, traditionally the correlated residuals are added back to the \texttt{CLEAN} image, thereby making it neccessary to estimate the covariance matrix of the residuals. 
Also, it is well known that \texttt{CLEAN} has some specific issues in deconvolution that are the result of the model approximation used \citep[i.e. the number of delta functions][]{fomalont}.\footnote{E.g. \texttt{CLEAN} faces difficulties representing the extended structure because the model consists of point sources ($\delta$-functions).}. As shown in the simulations of \cite{2001ApJ...554..948L} \texttt{CLEAN} creates a correlated error in the image plane. Thus the uncertainty in the image plane is non-uniformly distributed. Attempts to empirically account for the \texttt{CLEAN}-specific error were made by \cite{mojave8rm}. By using simulations the authors measured an additional variance in some points of the artificial source which they have attributed to \texttt{CLEAN}-specific errors. A formula to account for this additional source of $rms$ was introduced:
\begin{equation}
\label{eq:hovattaformula}
    \sigma = \sqrt{\sigma_{\rm rms}^2 + \sigma_{D_{\rm term}}^2 + (1.5\cdot\sigma_{\rm rms})^2} 
\end{equation}
where $\sigma_{\rm rms}$ is the $rms$ of pixel flux calculated within some area outside of the \texttt{CLEAN}-region, $\sigma_{D_{\rm term}}$  accounts for the instrumental polarization error contribution\footnote{It is shown to be significant for linear polarization (i.e. for Stokes parameters $Q$ and $U$, but for stokes $I$ it should be small in most cases).}, and the last term accounts for the inflated variance of pixel flux observed in simulations which is attributed to a \texttt{CLEAN}-specific error. This approach is now conventionally used to estimate the $rms$ of the flux of the \texttt{CLEAN} image pixels\footnote{We will subsequently call this \textit{pixel rms}.}. Obviously, this relation does not account for the non-uniformity of the error across the source. Also because it was derived as an empirical formula based on considering \texttt{CLEAN}-models for a couple of sources, it does not account for possible dependence on the individual source structure.

\subsection{Quantifying the uncertainty of combinations of images}

\subsubsection*{Spectral index}
Multifrequency VLBI observations of AGN jets provide important information on the physics of the outflows. Maps of the spectral index $\alpha$, where the spectral flux $S_\nu$ observed at the frequency $\nu$ is $S_\nu \propto \nu^{\alpha}$, can be used to constrain the nature of the jet components \citep[e.g.][]{mojave_spectral} or the VLBI core \citep{lisakov2017}, jet magnetic field configuration and acceleration mechanism of the emitting particles \citep{2011MNRAS.415.2081C}. When calculating the uncertainties of the spectral index distribution the correlated noise in the image plane is generally ignored \citep{fromm}. However it could be the source of fake patterns in the spectral index distribution and thus should be taken into account.

\subsubsection*{Significance of transverse Rotation Measure gradients}
Transverse RM gradients were predicted for jets carrying a helical magnetic field \citep{blandford1993} and successfully discovered by \cite{asada} in source 3C~273. Subsequent works \citep{contopoulos,gabuzda2004} claimed detection of many RM gradients. The failure to detect RM gradients in even well resolved quasar jets and the results of simulations with physical models of the helical magnetic fields \citep{broderick} resulted in \cite{taylor_3} proposing a set of criteria for the significance of the RM gradient. The most stringent requirement put forward was that a significant RM gradient should be detected with a length not smaller than three ``resolution elements'' (usually considered as three beam widths).

It was noted in \cite{mahmud_towers_2013} that the criteria proposed by \cite{taylor_3} are too conservative and they, therefore, increase the significance of the RM gradient detections by decreasing the statistical power (i.e. probability to detect the true RM gradient).
The influence of noise on the significance of the observed RM gradients was investigated in \cite{mojave8rm}. Simulations with artificial sources with $RM=0$ (i.e. unaffected by Faraday Rotation) allowed \cite{mojave8rm} to estimate the False Positives Rates (FPR) for detecting RM gradients with their observational setup. They concluded that with FPR nearly equal to $1\%$ one can use criteria such as two beam widths as a minimal gradient length and 3 $\sigma$ as a minimal change of RM value, where $\sigma$ is the mean RM error at the endpoints of a transverse slice of a jet.

However, the proposed criteria could be criticised in several ways. First, a non-local hypothesis test of the RM gradient significance depends only on the two endpoints of the corresponding RM slice. It is desirable for the test to account for all the data. Second, the endpoints used in the test depend on the chosen clipping level of a polarized flux. Finally,  further we show that the $rms$--based approach to estimating the error in the pixel flux results in statistically non-optimal estimates.   

Moreover the proposed criteria is only applicable to the distribution of the polarized flux used in simulations. This \textit{null hypothesis} model consists of the distribution of Stokes parameters $Q$ and $U$ equal to some constant fraction of the Stokes $I$ distribution. The real distribution of the polarized flux is actually unknown due to the uncertain RM contribution, noise, and convolution with the beam.
While it seems natural to assume a constant RM distribution in the simulations that estimate FPR, the possible dependence of the RM estimates on the true (i.e. at  infinite frequency) polarized intensity could result in an unaccounted bias in the FPR estimate. To partially overcome this one should use the null model of the polarized flux that is equal to the estimated one (i.e corrected for the measured RM).

With these caveats in mind it is desirable to formulate a well justified procedure for estimating the significance of the RM gradients that relies only on the data at hand, uses them all, and is free of subjective criteria. In Section~\ref{sec:rmgrad} we propose a criterion for statistically significant RM gradient that addresses these issues.

\section{Method}
\label{sec:method}

Bootstrap \citep{efron} is a kind of resampling method \citep{2012msma.book.....F} that is used mainly for assessing the accuracy of estimates. It is a robust alternative to the asymptotic parametric methods when their assumptions could be violated or when it is difficult or impossible to obtain a closed-form analytical expression for uncertainty estimates. The main idea of bootstrap is to approximate the process of data generation and to explore the sampling properties of the estimate of interest using a large number of artificially generated data sets.

We follow \cite{intro} in describing bootstrap and different bootstrapping methods. In general one observes some data $\mathbf{y}$, ($y_i$, i=$\overline{1, N}$) -- the vector of the measured data values -- and estimates some \textit{statistic} (that is a function of the data) of interest $s(\mathbf{y})$ and its associated uncertainty. There is an underlying probability mechanism $P$ (or data generating process, DGP) that produces the measured data, e.g. $\mathbf{y} = f(\mathbf{x}, \mathbf{\theta}) + \mathbf{\epsilon}$, where $f(\mathbf{x}, \mathbf{\theta})$ is some underlying unknown model with a vector of covariates $\mathbf{x}$, e.g. points where data are measured, $\mathbf{\theta}$ is a vector of model parameters and  $\mathbf{\epsilon}$ is the model of noise (e.g. independent and identically distributed Gaussian noise). One estimates $s(\mathbf{y})$ using e.g. best fit values of the model parameters: $\hat{\mathbf{\theta}} = s(\mathbf{y})$. In general the algorithm used to estimate $s(\mathbf{y})$ could be some highly nonlinear algorithm or even a computer algorithm \citep{chernik}.

Bootstrap allows one to estimate the error of $s(\mathbf{y})$ by approximating the data-generating probability mechanism $P$ and simulating the observed data many ($B$) times with the approximated DGP, each time obtaining the so called \textit{bootstrap sample} $\mathbf{y}_{\mathrm{boot}}$. Then each of the generated bootstrapped samples is used to estimate the desired statistic $s(\mathbf{y}_{\mathrm{boot}})$. Their distribution is used to estimate the uncertainty of the statistic $s$ for the original (i.e. the observed) sample $s(\mathbf{y})$.

Probability mechanism $P$ could be approximated in different ways. One can assume that $P=(F_{y})$, where $F_{y}$ is the probability distribution function of data. Because one does not know $F_{y}$ , we can approximate it (and thus $P$) by the empirical distribution function $F_{y_{\mathrm{obs}}}$:
\begin{equation}
\hat{P} = (\hat{F}_{y_{\mathrm{obs}}})
\label{eq:1}
\end{equation}

If one has a model fitted to the data e.g. $\mathbf{y} = f(\mathbf{x}, \mathbf{\theta}) + \mathbf{\epsilon}$, then the probability model can be written as $P = (\mathbf{\theta}, F_{\epsilon})$, where $F_{\epsilon}$ is the noise probability distribution of the regression model. In that case $P$ can be approximated as:
\begin{equation}
\hat{P}=(\hat{ \mathbf{\theta}}, \hat{F}_{\epsilon})
\label{eq:2}
\end{equation}
where $\hat{ \mathbf{\theta}}$ is the estimate of the model parameters and $\hat{F}_{\epsilon}$ is the empirical distribution function of the residuals between the data and fitted model. E.g., if the model of the noise (e.g. i.i.d. Gaussian noise) and its variance $\hat{\sigma}^2$ are known, then:
\begin{equation}
\hat{P}=(\hat{ \mathbf{\theta}}, F^{\mathrm{norm}}(\hat{\sigma}^2))
\label{eq:3}
\end{equation}
where $F^{\mathrm{norm}}$ is the Gaussian distribution with the estimated variance $\hat{\sigma}^2$.

Thus bootstrap consists of the two main approximations: approximation of the probability mechanism $P$; and approximation due to the finite number $B$ of bootstrapped samples. The latter is $\propto B^{-0.5}$ and can be made negligible relative to the former \citep{intro}. Depending on the approximation used bootstrap is called \textit{non-parametric} or ``\textit{pairs}'' bootstrap if (\ref{eq:1}) is used (originally introduced by \citealt{efron}), \textit{residuals} bootstrap if (\ref{eq:2}) is used, and \textit{parametric} bootstrap if (\ref{eq:3}) is used. These definitions reflect the way bootstrap samples are generated under the selected approximation of the data generating process. In the pairs bootstrap the whole observations $\mathbf{y}_{\mathrm{obs}}$ are resampled from the corresponding $F_{y_{\mathrm{obs}}}$, where $\mathbf{y}_{\mathrm{obs}}$ can include covariates in the regression problem. Because each of the $N$ observed data points has a probability of $1/N$, resampling occurs by sampling with replacement from the original data set $N$ times. The pairs bootstrap is the most robust to model misspecification because its approximation of the data generating process (\ref{eq:1}) does not make any assumptions about the true model.

In the residuals bootstrap the data set is replicated by sampling with replacement from the empirical distribution of the residuals between the observed data and fitted model and then followed by adding the obtained residuals to the model predictions. Residuals bootstrap assumes that $F_{\epsilon}$ does not depend on covariates i.e. the residuals are homoscedastic. But model $P$ could be modified to account for the heteroscedasticity of the residuals, i.e. their dependence on covariates \citep{intro,davison_hinkley_1997_6}. The residuals bootstrap is more sensitive to a model specification error than the pairs bootstrap. Nevertheless the model does not have to fit the observed data perfectly to give a reasonable result \citep{intro}. But if it is a good approximation then the residuals bootstrap delivers more efficient uncertainty estimates than the pairs bootstrap.

Conventional methods like least-squares fitting basically make the same assumptions as the parametric bootstrap\footnote{In general the parametric bootstrap agrees with the maximum likelihood \citep{esl}.}, but some methods does not have a closed-form expression for error estimates (e.g. nonlinear regression or some sophisticated computer algorithm). If the model approximates DGP well then the parametric bootstrap delivers the most efficient error estimates. According to the ``bias-variance trade-off'' this comes at the cost of the biased error estimates when the model used is a bad approximation to the true model underlying the DGP (see Section~\ref{sec:limits})).

\subsection{Bootstrap for VLBI data}
\label{sec:boot4vlbi}
In the case of VLBI observational data  $\mathbf{y} = \mathbf{V}_{\mathrm{obs}} = (V_{\mathrm{obs},i}),$ $i=\overline{1, N_{\mathrm{obs}}}$, where $\mathbf{V}_{\mathrm{obs}}$ are $N_{\mathrm{obs}}$ measurements of the visibility function at certain points $\mathbf{x} = (\mathbf{u}, \mathbf{v}) = (u_i, v_i)$ of the $(u,v)$-plane. The uncertainty of some visibility data statistic $s(\mathbf{V}_{\mathrm{obs}})$ is estimated using traditional methods, i.e. image deconvolution, direct visibility modelling, where $s$ could be the flux of certain pixel(s) in the \texttt{CLEAN} image, parameters of a simple model fitted directly to the interferometric visibilities $\mathbf{V}_{\mathrm{obs}}$ etc. Thus, in general, certain function with parameters $\mathbf{\theta}$ is ``fitted'' to the observed visibilities $\mathbf{V}_{\mathrm{obs}}$ by some algorithm. 
To bootstrap the visibility data one has to decide which approximation from (\ref{eq:1})--(\ref{eq:3}) to use.
Resampling visibilities with their $(u,v)$-points, i.e. using approximation (\ref{eq:1}), could result in biased
estimates due to the degraded resolution. It is expected that $e^{-1}$ ($36.8$ \%) of data points will not get into a single bootstrapped data set \citep{esl}. Thus the resampling procedure that corresponds to (\ref{eq:1}) effectively reduces the coverage of the $(u,v)$-plane in every single bootstrapped sample.

To avoid the problem of different coverage of the $(u,v)$-plane one can use approximation (\ref{eq:2}) and resample the residuals between the observed and model visibilities. This resampling scheme keeps the $(u,v)$-coverage for each of the bootstrapped samples \citep[and, thus, its informational content,][]{davison_hinkley_1997_6} the same as for the original data set. Thus $P$ in VLBI observations could be approximated by $\hat{P} = (\hat{\mathbf{\theta}}, \hat{F}_{\epsilon})$, with $\mathbf{V}_{\mathrm{obs}} = f((\mathbf{u}, \mathbf{v}), \mathbf{\theta}) + \mathbf{\epsilon}$, where $f((\mathbf{u}, \mathbf{v}), \mathbf{\theta})$ is the model fitted by some algorithm\footnote{The model can include instrumental parameters, {e.g. antenna gains} -- not only the parameters that determine the source brightness distribution.} (e.g. deconvolution algorithm), and $\hat{\mathbf{\theta}}$ are the estimated parameters of the model (e.g. list of \texttt{CLEAN} components for \texttt{CLEAN}-deconvolution, set of Gaussians for direct fitting of the visibility measurements). $\hat{F}_{\epsilon}$ is the empirical distribution of the residuals between the observed visibilities and model values. If $\hat{F}_{\epsilon}$ may be approximated by some parametric form, e.g. Gaussian, one can go further and assume a fully parametric approximation (\ref{eq:3}).

As already noted, (\ref{eq:2}) requires that the residuals distribution should not depend on covariates $(u_i, v_i)$. For VLBI observations different baselines of the interferometer may have different sensitivities and this is especially true for Space VLBI. Thus one should alter the model of $P$ to include the distribution of the residuals on different baselines individually $\hat{P} = (\hat{\mathbf{\theta}}, \hat{F}_{\epsilon, j})$, where $j=\overline{1, N_{\mathrm{baselines}}}$ is the baseline index. This effectively assumes a different value of noise on different baselines. Baseline-independent resampling of residuals could bias the resulting uncertainty estimates in case of a single ``noisy'' baseline that supplies residuals to other more sensitive baselines in the bootstrapped data sets.

\subsection{Implementation}
Taking into account the results of Section~\ref{sec:boot4vlbi},
the algorithm of the error estimation for certain visibility data statistic can be specified as follows (see Algorithm~\ref{algo1} for schematic representation). First, one obtain self-calibrated visibilities $V_{\mathrm{sc}}$ and parameters $\hat{\theta}$ of the model fitted to visibilities. The estimate of interest is often some function of the model parameters (e.g. flux of a pixel(s) in the \texttt{CLEAN} image or position of a single Gaussian component). Then one calculates the residuals $V_{\mathrm{res}}$ between the observed and model visibilities. For each baseline the residuals are optionally adjusted, e.g. filtered from outliers and fit with some probability density model (see Section~\ref{sec:modifications}). Then the bootstrapped data set is created $B$ times by sampling with replacements from the residuals (or from their density estimate) on each baseline and adding the obtained resamples to the model visibilities. The bootstrap data sets $V_{\mathrm{boot}, i}$  are analyzed exactly in the same way as the observed one. The obtained distribution of the statistic of interest $s(V_{\mathrm{sc}})$ is used to estimate its uncertainty $\hat{\sigma}_{s}$.

\begin{algorithm}
\caption{Bootstrapping self-calibrated visibility data $V_{\mathrm{sc}}$ using model $V_{\mathrm{model}}$ (with parameters $\theta$) to estimate the uncertainty of some statistic $s(V_{\mathrm{sc}})$}\label{boot1}
\begin{algorithmic}
\State $\hat{\theta} \gets \texttt{"fit" model $V_{\mathrm{model}}(\theta)$ to } \textit{$V_{\mathrm{sc}}$}$
\State $\textit{$V_{\mathrm{res}}$} \gets \textit{$V_{\mathrm{sc}}$} - \textit{$V_{\mathrm{model}}(\hat{\theta})$}$
\State $\texttt{Optional adjustment of }$ $\textit{$V_{\mathrm{res}}$}$
\For{ $i$ \texttt{ from} 1 \texttt{to} $B$}
\For{ $j$ \texttt{ from} 1 \texttt{to} $N_{\mathrm{baselines}}$}
    \State $\textit{$V_{\mathrm{res}, i}^j$} \gets \texttt{sample with replacement from } \textit{$V_{\mathrm{res}}^j$}$
\EndFor
    \State $\textit{$V_{\mathrm{boot}, i}$} \gets \textit{$V_{\mathrm{model}}$} + \textit{$V_{\mathrm{res}, i}$}$
    \State $\textit{$\hat{\theta}_{\mathrm{boot}, i}$} \gets \texttt{fit model to } \textit{$V_{\mathrm{boot}, i}$}$
    \State $\textit{$\hat{s}_{\mathrm{boot}, i}$} \gets \texttt{estimate from } \textit{$\hat{\theta}_{\mathrm{boot}, i}$}$
\EndFor

\State  $\texttt{Estimate error}$ $\hat{\sigma}_{s} \gets \textit{$\hat{s}_{\mathrm{boot}, i}$}$
\end{algorithmic}
\label{algo1}
\end{algorithm}

\subsection{Modifications}
\label{sec:modifications}
If the distribution of the residuals between the observed and model data can be approximated by some parametric density (e.g. Gaussian or t-distribution), then one can repeatedly draw samples from the fitted distribution instead of directly sampling from the raw residuals. This effectively uses the parametric bootstrap (\ref{eq:3}). One can also fit some nonparametric model to the residuals distribution on each baseline, e.g. kernel density estimate (KDE) or Gaussian mixture model \citep{esl}. Finally, one can just use raw residuals and resample from them. The advantage of using the more constrained model of the residuals is that it results in more effective error estimates in case of a good approximation of the underlying DGP. Conversely, if the model is clearly wrong, using the parametric bootstrap introduces a bias to the resulting errors. As discussed above the resampling step should be done on the per baseline basis, especially if the sensitivities of the individual baselines are different. Heteroskedastic residuals within the individual baselines could be treated using the ``wild'' bootstrap \citep{wu}.

Our implementation consists of the following steps. First, we search for outliers on each baseline, frequency band, and visibility hands\footnote{parallel $\langle RR^* \rangle $, $\langle LL^* \rangle $ and cross hand correlations $\langle RL^* \rangle $, $\langle LR^* \rangle $, where $R$ and $L$ denote voltages from left and right circular polarized feeds, brackets means averaging and star denotes complex conjugate \citep{2017isra.book.....T}.} separately using the DBSCAN algorithm \citep{dbscan}. We treat 1D residuals of the real and imaginary parts of self-calibrated visibilities as well as 2D residuals on the complex plane as features for a DBSCAN algorithm. DBSCAN is a density-based clustering algorithm that marks points as outliers if they are isolated in low-density regions. The KDE of the outlier-filtered residuals distributions for real and imaginary parts are obtained\footnote{\textit{Scikit-learn} Python package \citep{scikit-learn} was used for these steps.}.
The width of the KDE Gaussian kernel is found using 5-fold Cross-Validation \citep{esl}. Only positively weighted visibility measurements were used during this procedure.

As already noted, this scheme could be extended to account for instrumental effects, e.g. self-calibration errors. In that case the model $f$ underlying DGP will include gains of the antennas as parameters. In a similar manner the influence of any other instrumental effect (e.g. polarization leakage or overall amplitude calibration offset) on the uncertainty estimate can be accounted for if its value can be estimated.

\section{Comparing bootstrap-based errors to the conventional errors by using simulations}
\label{sec:simulations}

\subsection{Coverage}
\label{sec:coverage}

Recall that by definition the \textit{confidence interval} (CI) is a region that contains the ``true'' value of the parameter of interest (e.g. flux of a pixel) with a specified frequency or \textit{confidence level} \citep{Wasserman:2010:SCC:1965575}, say 95$\%$.
This means that the confidence interval is supposed to contain the true value of the parameter with frequency equal to 0.95 under the repeated constructions of CI using independent observational data sets. But if some assumptions underlying the procedure used to build CI are violated, e.g. data sets are not normally distributed for asymptotic CI of the mean, then the fraction of CIs that contain true values could differ from the nominal value that is described by the confidence level (0.95 in this case). The \textit{coverage probability} or \textit{coverage} of the interval estimate of a parameter (e.g. confidence interval) is the long-run probability for this estimate to contain the true value of the parameter. For VLBI observations the coverage of the CI for the pixel flux could be represented as follows. The source is observed with VLBI and the resulting data set is calibrated and imaged with the \texttt{CLEAN} algorithm. By using a certain procedure CI for the pixels, the flux is obtained. Then the same source is observed again many times with the same VLBI array (with the same sensitivity, time interval, and ($u$, $v$)-plane coverage). The source is assumed to be stationary. Then the coverage of the CI for the pixels' flux is the fraction of cases where those CIs contain true values. The same applies for any other function of the observed data, e.g. the size of the fitted Gaussian component or the shift between images obtained at two frequencies. As the definition of coverage involves the notion of the true value of the parameter, simulations with a known model are required to assess the coverage.

Optionally $1 - \alpha$-confidence interval (that is, CI with $1 - \alpha$ confidence level) should have coverage $1 - \alpha$, where $\alpha$ is the \textit{significance} of the corresponding hypothesis test, i.e. Type I error rate. Actual coverage could depart from the nominal value if some assumptions that are used in the error estimation procedure are broken. If the coverage is less then the nominal one, then the statistical significance of the results based on such confidence regions will be overestimated. This situation could result from biased estimates or from errors that are too small.
Otherwise, if the coverage is higher than the nominal value, the errors are overestimated and the statistical power of the corresponding hypothesis test (with the null hypothesis being that parameter equals some specified value) decreases.

\subsection{Stokes I intensity images}
\label{sec:icoverage}
\begin{figure}
   \centering
   \includegraphics[width=\columnwidth, trim=0.3cm 0.9cm 0.3cm 0.3cm]{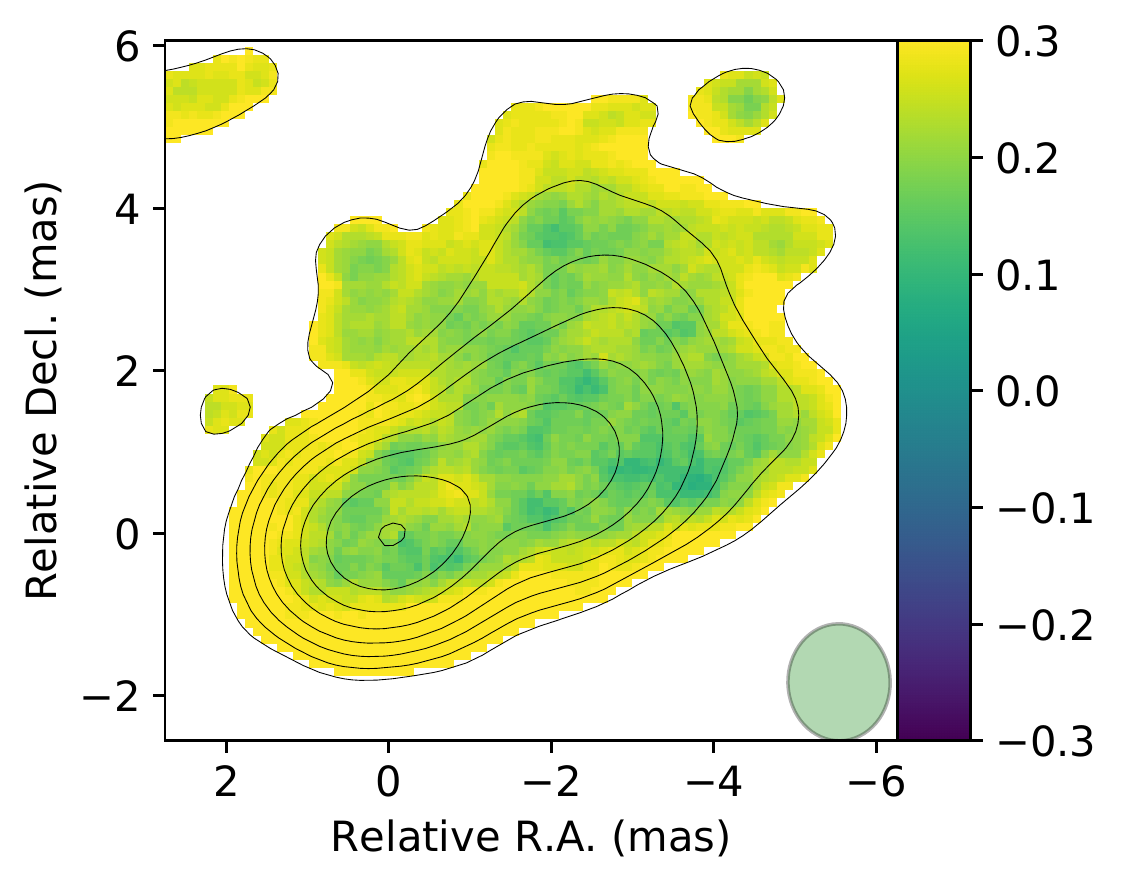}
      \caption{Image of deviations from the nominal coverage of $rms$-based 68$\%$ confidence interval for stokes I for 1749$+$701. The green ellipse in the image show the synthesized beam.}
         \label{1749_cov_conv}
   \end{figure}

\begin{figure}
   \centering
   \includegraphics[width=\columnwidth, trim=0.3cm 0.9cm 0.3cm 0.3cm]{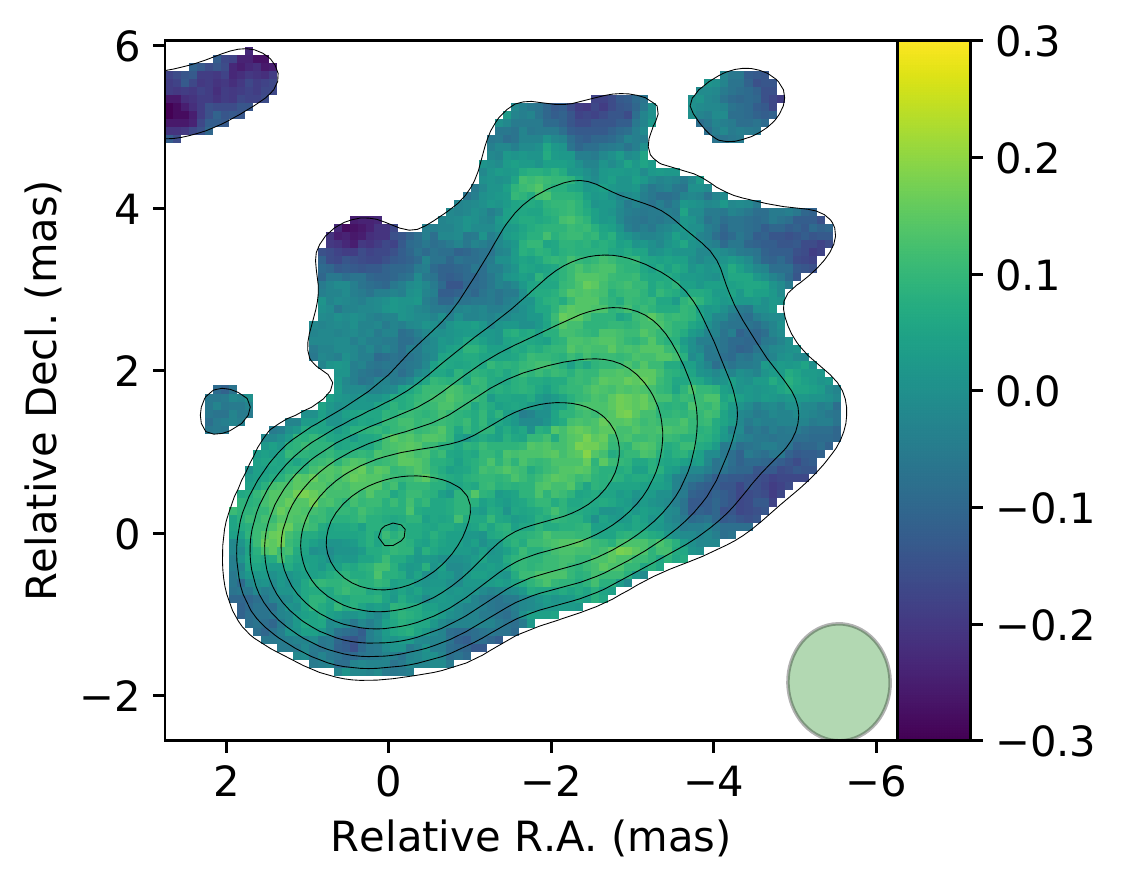}
      \caption{Image of deviations from the nominal coverage of bootstrap-based 68$\%$ confidence interval for stokes I for 1749$+$701.}
         \label{1749_cov_boot}
   \end{figure}

\begin{figure}
   \centering
   \includegraphics[width=6cm, trim=0.3cm 0.5cm 0.3cm 0.3cm]{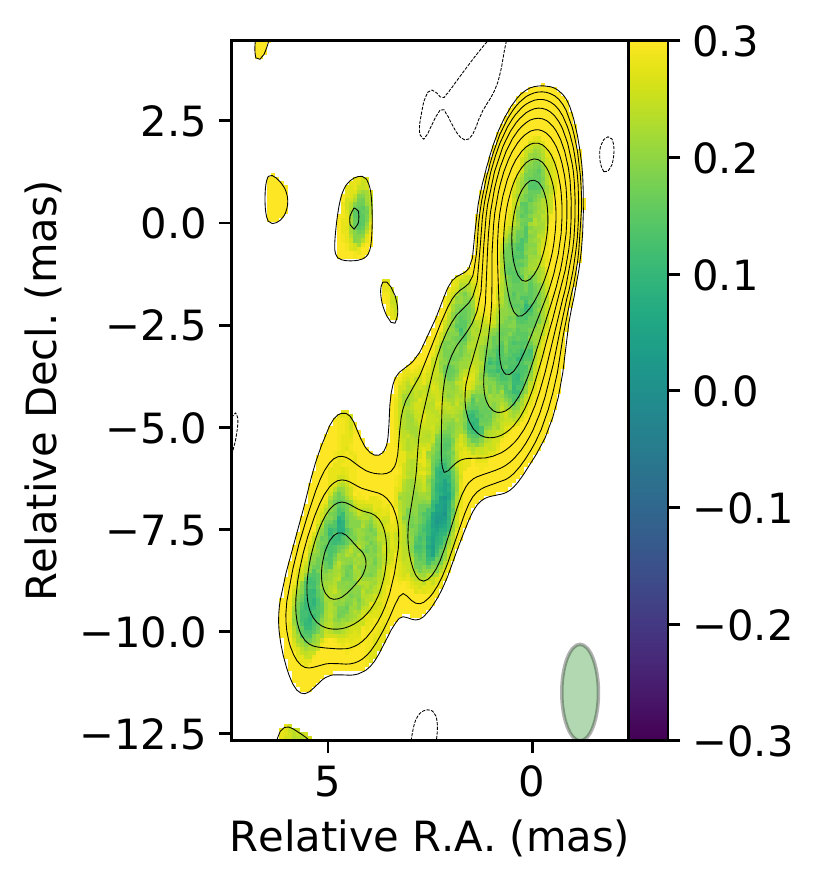}
      \caption{Image of deviations from the nominal coverage of $rms$-based 68$\%$ confidence interval
       for stokes $I$ for 1514$-$241.}
         \label{1514_cov_conv}
   \end{figure}

\begin{figure}
   \centering
   \includegraphics[width=6cm, trim=0.3cm 0.5cm 0.3cm 0.3cm]{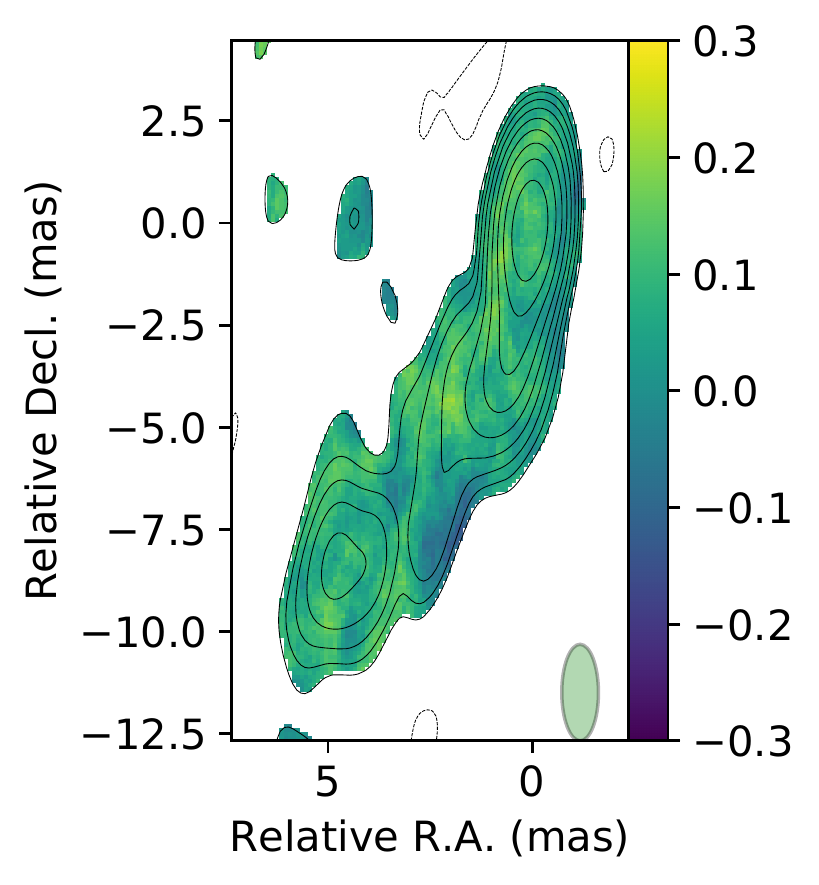}
      \caption{Image of deviations from the nominal coverage of bootstrap 68$\%$ confidence interval
       for stokes $I$ for 1514$-$241.}
         \label{1514_cov_boot}
   \end{figure}

In order to compare the coverages of the CI for the \texttt{CLEAN} image pixel flux constructed using bootstrap and conventional ($rms$-based) approaches we conducted simulations with a known model. We selected two AGN radiosources with different declinations to compare the results for different $(u,v)$-coverages: $1749+701$ and $1514-241$. We fetched self-calibrated data from the MOJAVE \citep{mojave6kinematicsanalysis} database (8.1 GHz X-band and epochs 2006-04-05 and 2006-04-28) and used the corresponding \texttt{CLEAN} models as ``ground truth'' models. Then we created a sample of 100 simulated observations by adding noise to model visibilities with the value estimated on each baseline of the original data sets. Finally we deconvolved the resulting artificial data sets using \texttt{CLEAN} and obtained two maps of the coverage of the confidence intervals for the pixels' flux:  one for bootstrap-based errors and one for conventional $rms$-based errors. We followed the procedure described in Section~\ref{sec:modifications} using $B$=100 bootstrap replications for each data set of the simulated sample to obtain bootstrap-based errors. The resulting maps of the coverage for 68$\%$ confidence intervals are shown for source 1749$+$701 (Fig.~\ref{1749_cov_conv} and Fig.~\ref{1749_cov_boot} for $rms$- and bootstrap-based errors) and source 1514$-$241 (Fig.~\ref{1514_cov_conv} and Fig.~\ref{1514_cov_boot}). Colours show the deviation of the coverage from the nominal value (0.68 in this case). A positive value means that coverage is higher than the nominal one. Two facts are apparent. First, the coverage is nonuniform especially for $rms$-based errors. Second, $rms$-based errors are substantially over-covered relative to the nominal value. Bootstrap-based  errors are closer to the nominal coverage and more uniform, especially in the outer parts of the source structure with relatively low signal-to-noise ratio (SNR).

\subsection{Rotation Measure maps}
\label{sec:rm}

Extending the bootsrapping approach to the problem of estimating the uncertainty of a non-linear combinations of images is straightforward.

\begin{figure}
   \centering
   \includegraphics[width=\columnwidth,trim=0.3cm 0.5cm 0.3cm 0.3cm]{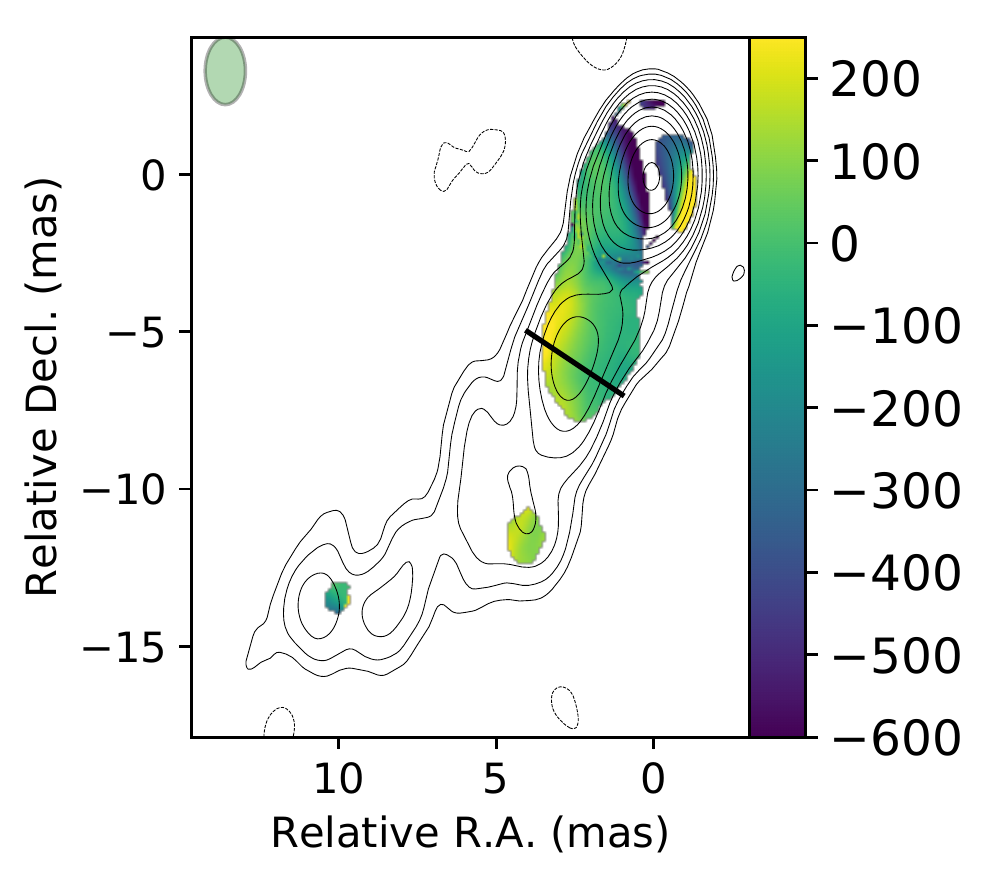}
      \caption{Rotation measure (RM) map overlaid on Stokes I contours for source 2230$+$314. The green ellipse in the image show the synthesized beam. Colour bar is in $\rm rad/m^{-2}$.}
         \label{rotmmap}
   \end{figure}

\begin{figure}
   \centering
   \includegraphics[width=\columnwidth,trim=0.3cm 0.5cm 0.3cm 0.3cm]{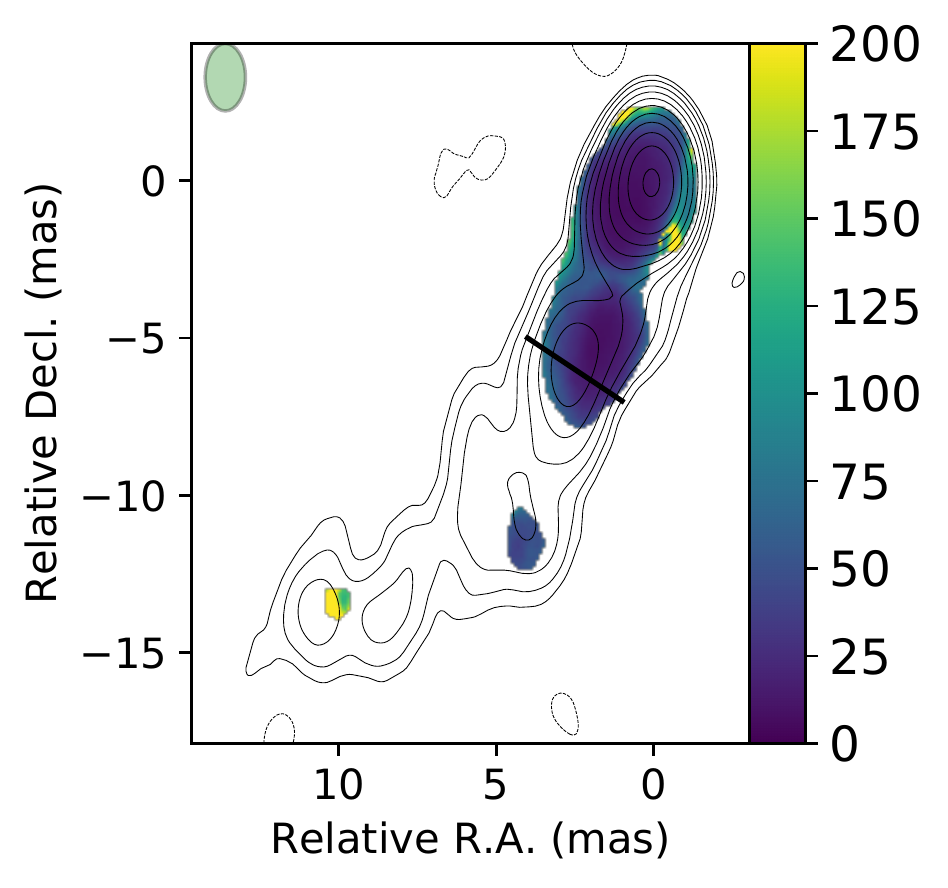}
      \caption{Bootstrap error distribution for RM map presented in Fig.~\ref{rotmmap}.}
         \label{rotmmaperrorboot}
   \end{figure}

\begin{figure}
   \centering
   \includegraphics[width=\columnwidth,trim=0.3cm 0.5cm 0.3cm 0.3cm]{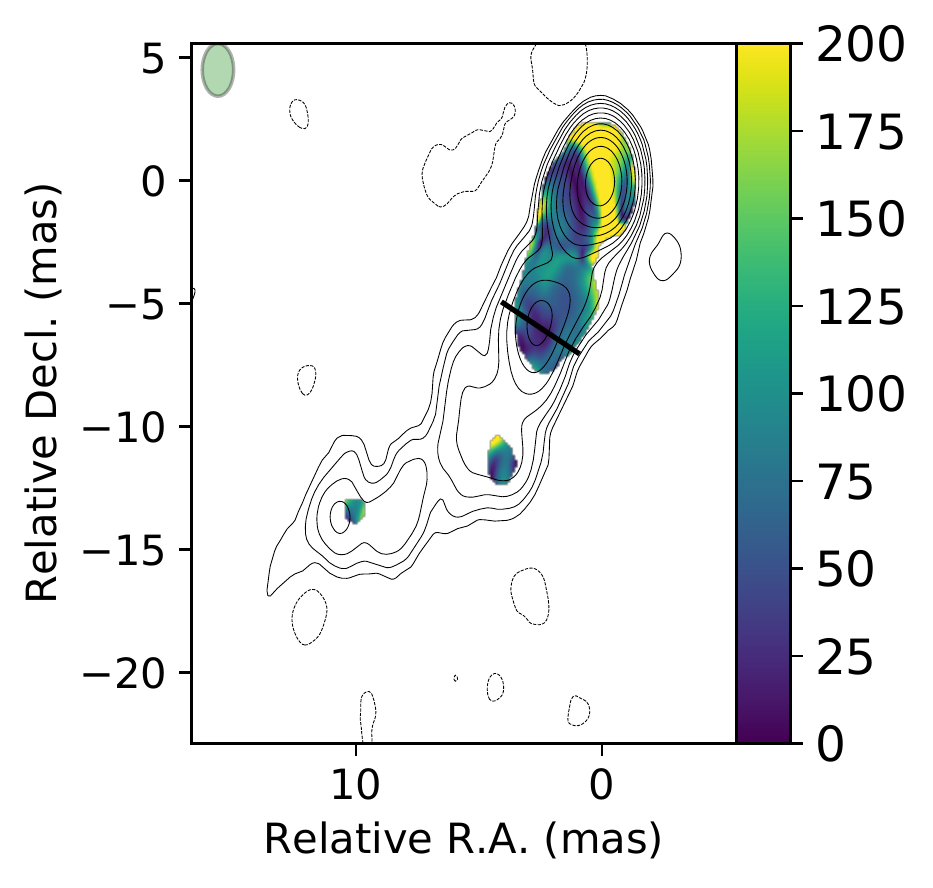}
      \caption{Conventional error distribution for the RM map presented in Fig.~\ref{rotmmap}.}
         \label{rotmmaperrorconv}
   \end{figure}

In Fig.~\ref{rotmmap} the rotation measure (RM) map of the source 2230$+$314 and in Fig.~\ref{rotmmaperrorboot} the map of the 68\% CI of RM obtained using bootstrap are shown. Errors given by the asymptotic covariance matrix \cite{mojave8rm} are presented in Fig.~\ref{rotmmaperrorconv}. Contribution from the absolute electric vector position angle (EVPA) calibration error is not included to enhance the difference between both approaches.

When comparing bootstrap-based (Figure~\ref{rotmmaperrorboot}) and conventional ($rms$-based, Figure~\ref{rotmmaperrorconv}) errors the most striking difference appears in the core region. While the former are minimal here (nearly 10 $\rm rad/\rm m^2$), $rms$-based errors of up to 150 $\rm rad/\rm m^2$ are observed in the core, which is nearly two times larger than the error in the low-SNR jet region. This counter-intuitive fact could be understood as the result of the broken linear dependence between the linear polarization positional angle (PPA) and the squared wavelength in the core region. Indeed, part of this region is masked by the procedure that compares $\chi^2$ values with the critical value for the given number of degrees-of-freedom as described in \cite{mojave8rm}. That means that the linear model used in the fit is a poor approximation of the observed $\lambda^2$-dependence. Asymptotic errors from the covariance matrix are useless in a situation with evident model-specification bias because they cannot estimate neither the variance (the method's assumptions are violated) nor the bias (one must know the true model to estimate the bias).

\begin{figure}
   \centering
   \includegraphics[width=\columnwidth, trim=0.3cm 0.5cm 0.3cm 0.3cm]{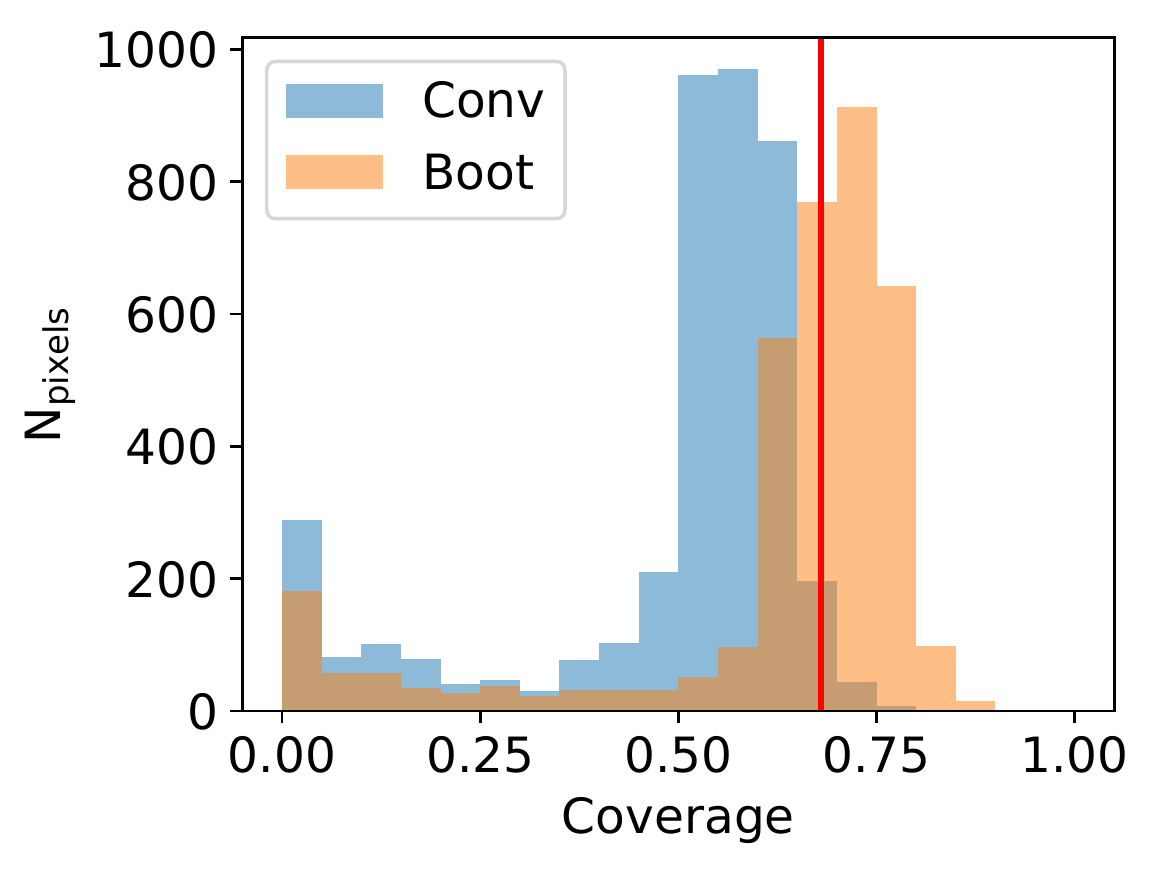}
      \caption{Histogram of pixels coverage distribution for the RM errors obtained using simulations described in Section~\ref{sec:rm}. Both bootstrap-based errors and conventional errors are shown. Red vertical line shows the nominal coverage.}
         \label{fig:rmcov}
   \end{figure}

We also conducted calibration tests to compare the coverages of the conventional and bootstrap-based errors. We used the model of the inhomogeneous jet from \cite{bk_jet} and MOJAVE \citep{mojave6kinematicsanalysis} real data at 8.1, 8.4, 12.1 and 15.4 GHz for source 1458+718 at epoch 2006/09/06. We transformed model images to the ($u$, $v$)-plane at the observed ($u$, $v$)-points and added noise estimated from real data sets at each frequency band. We used a constant fraction (0.2) of the Stokes $I$ image as a model for Stokes $Q$ and $U$. Using different realizations of noise we created a sample of artificial sources. For each source the RM map as well as its uncertainty map were obtained using the approach\footnote{We have not taken such instrumental effects as D-terms and EVPA-calibration uncertainty into account because we are interested in the influence of the correlated noise. Nevertheless, our analysis can be easily extended as discussed in Section~\ref{sec:modifications}.} of \cite{mojave8rm}. We also calculated the uncertainty maps using $B$=100 bootstrap replications of each artificial data set by resampling the residuals in the ($u$, $v$)-plane (as described in Section~\ref{sec:modifications}). Then coverage maps were created for both types of the uncertainties and the true model. The last one was obtained from the model distribution of \cite{bk_jet} in the same way as our artificial sample but with the noise added that equals a small fraction (0.01) of the observed value. The histograms of the coverage across the source are presented in Figure~\ref{fig:rmcov}. The red line shows the nominal coverage (0.68). It is apparent that bootstrap-based errors have coverage closer to the nominal one and $rms$-based errors are under-covered.

\subsection{Significance of the RM gradients}
\label{sec:rmgrad}

\begin{figure}
  \centering
  \includegraphics[width=\columnwidth,trim=0.3cm 0.5cm 0.3cm 0.3cm]{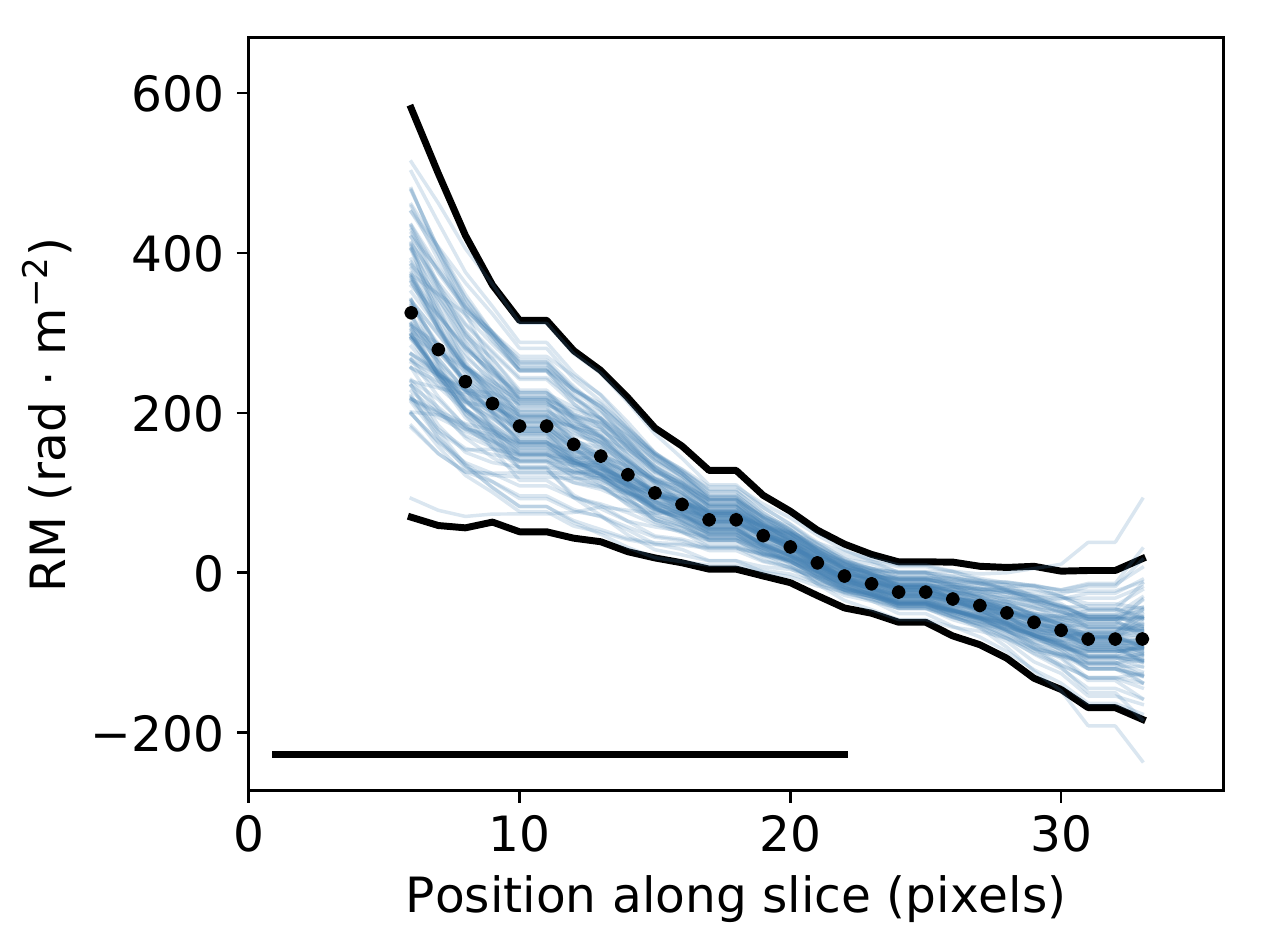}
     \caption{Measured values of RM along the transverse slice of 2230$+$114 jet (shown as black line in Figure~\ref{rotmmap}) with simultaneous confidence band made from 100 bootstrap realizations. The straight black line in slice show the synthesized beam}
         \label{fig:rotm_slice}
  \end{figure}

The significance of the RM gradient could easily be tested statistically once reliable estimates of the uncertainties are obtained. Recall the duality between confidence intervals and hypothesis testing \citep{lehmann2005testing}. Using bootstrap replications of the RM image one can build the so called \textit{simultaneous} confidence bands \citep[SCB][]{simcb} for the estimated RM profile in any slice of the RM map (e.g. Figure~\ref{fig:rotm_slice}). It is important to use simultaneous confidence bands to avoid the problem of multiple comparisons. This situation arises while testing several statistical hypothesis simultaneously and results in an overestimated significance.
Any RM curve that can be embraced by an SCB is consistent with the observed data at the selected level of significance $\alpha$. (e.g.  0.05 for $95\%$ CB).
Thus, we can reject the null hypothesis that the RM gradient is absent when the SCB fails to embrace at least one horizontal line\footnote{We note that this criterion should be considered only as an observational one. Other, i.e. physical criteria, should also be fulfilled (see e.g. \citealt{taylor_3}). Thus the proposed criterion is necessary but not sufficient.}. Here the horizontal line represents the absence of the RM gradient. In other words SCB must be pierced by at least one horizontal line from both sides for the corresponding RM gradient to be statistically significant. By using the simultaneous CB in the criterion we require that horizontal line will be embraced by the corresponding CIs in \textit{each} point of the slice. Thus the number of simultaneously tested hypotheses equals the number of slice points. Using pointwise CB consisting of the CIs of the individual slice points could then result in a situation where the horizontal line misses a particular CI simply by chance. With the slice consisting of $n$ points and a single point CI with $1-\alpha$ confidence level the probability of the occurrence of at least one such point is $1 - (1 - \alpha)^n$. For typical widths of a quasar jets $n \sim 10$. For a typical quasar jet width $n \sim 10$ and the conventional value of $1 - \alpha = 0.95$ this probability is $\sim 0.4$, which is much less then the specified confidence level (0.95).

Criteria for the significant RM gradient used in \cite{mojave8rm} depends on a user specified level at which PPA images are clipped before calculating the RM map. If clipping is done at a different level the errors at the slice edges will change accordingly and the criteria of \cite{mojave8rm} could give different results in low signal-to-noise jet regions\footnote{Quasar jets are typically linearly polarized at a level of 10$\%$ \citep{mojave_1}.}. Given the proposed criterion of a significant RM gradient one does not need to clip PPA images before producing RM maps. At the source-free part of the RM image the SCB of the RM profile will be wide because only noise contributes to the measured RM. However the criterion concerns the regions of the SCB that are steep and/or narrow enough to be pierced through by a horizontal line. Therefore it is free from the crude step of choosing the width of the transverse slice with a hypothetical RM gradient.
We would like to note that the proposed criterion does not depend on the number of resolution elements spanned by the transverse RM slice. Indeed, the effective resolution depends on the SNR and cannot be the same across the source.

\subsection{Spectral index maps}
\label{sec:spindex}

\begin{figure}
   \centering
   \includegraphics[width=\columnwidth, trim=0.3cm 0.5cm 0.3cm 0.3cm]{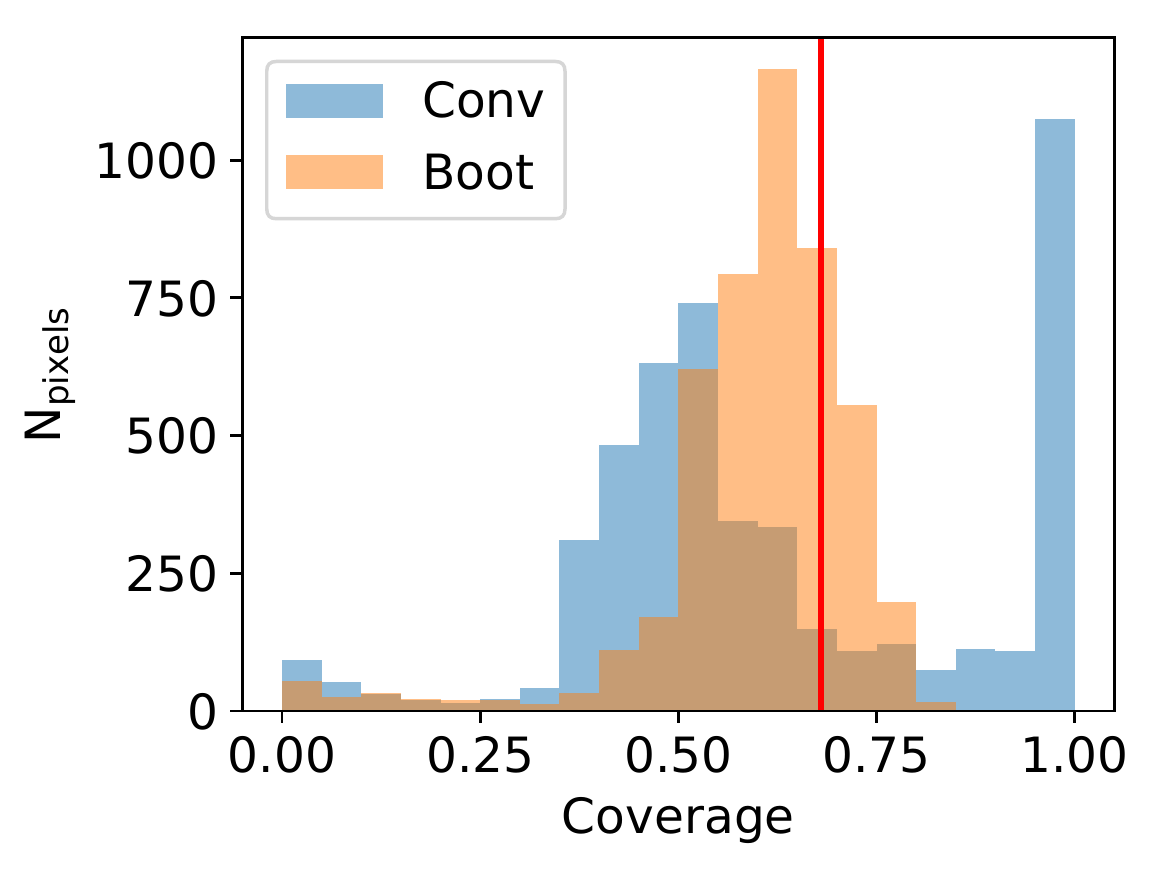}
      \caption{Histogram of pixel coverage distribution for spectral index errors obtained using simulations described in Section~\ref{sec:spindex}. Both bootstrap-based errors and conventional errors are shown. Red vertical line shows the nominal coverage.}
         \label{fig:spcov}
   \end{figure}

We conducted calibration tests for spectral index maps using the same procedure as for the Rotation Measure maps (see Section~\ref{sec:rm}). The spectral index in each pixel was calculated by fitting a single power-law to all four frequency bands. We calculated conventional errors following \cite{mojave_spectral} but without the contribution of the uncertainty of the amplitude calibration. The results are presented in Figure~\ref{fig:spcov}. The coverage of bootstrapped-based errors is closer to the nominal value and is more uniform.

\section{Discussion}
\label{sec:discussion}

\subsection{Possible limitations of applying bootstrap to VLBI}
\label{sec:limits}

With the assumptions and approximations implied by the method and presented in Section \ref{sec:method} we can identify the following cases in which applying bootstrap to VLBI data should be made with caution:

\begin{itemize}
\item \textit{Model is clearly inconsistent with data.}
When the model is clearly inconsistent with the data the distribution of the residuals in the $uv$-domain will depend on the $uv$-point inside a single baseline thus violating the assumption of their independence. Using the method blindly will then result in biased errors of the model parameters\footnote{It should be noted that in some sense \textit{pairs} bootstrap~(\ref{eq:1}) can be used as a proxy to estimate the model approximation error. Indeed, when the model is a bad approximation of the data the errors of the model parameters become larger. See also \citep{2016arXiv160805913K} for using bootstrap to estimate the model error.}. Thus one should check the model before using the method to be sure that the obtained uncertainty estimates are reliable. If the model is clearly inconsistent with the data the best one can do is to use the fully parametric bootstrap and resample not the modified residuals but the Gaussian noise with the variance estimated from the observed visibilities. One can use, e.g., the successive differences approach \citep{briggs} or Stokes V visibilities that are supposed to have a source signal at a level of only tenths of a percent for quasar jets \citep{vitr}.

\item \textit{Simple model fitted to complex brightness distributions.}
Using models that are too simple to account for the observed brightness distribution brings issues similar to those mentioned above. But we would like to distinguish this case in a separate item to highlight the consequences of fitting simplistic models to complex brightness distributions. The squared error of the model prediction can be decomposed into a sum of the variance (random or estimation error) and the bias (systematic or model approximation error) squared\footnote{There is also a contribution from the irreducible error due to noise.}. The variance shows how stable the model predictions are when one uses different data sets for fitting the model. The bias represents the error due to using a simplistic function to approximate the intrinsically more complex relation. A model that is clearly too simple to account for the observed brightness distribution will have significant bias but, at the same time, its variance will be small relative to the bias due to the ``bias-variance trade-off'' \citep{esl}. Bootstrap allows one to estimate the random errors of the model parameters, i.e. errors that are due to the model variance and not to the bias. Thus the model parameters could be highly constrained by the data (i.e. reveal relatively small estimated errors) even if the model does not represent the observed data well. This is especially pronounced in case of high SNR data sets. The value of SNR at which the bias begins to dominate in the total error depends on the model used and the size of the data set. ``Learning curves''\footnote{That is, the dependence of the prediction performance of the model on the amount of data used to fit it.} \citep{raschka2015python} could be used to quantify the complexity of the model and to ensure that the bias is not a dominant factor in the total error, otherwise the errors estimated using bootstrap will be underestimated. One can use e.g. the method described in \cite{2016arXiv160805913K} to estimate the model approximation error in such cases using bootstrap. We discuss this in the context of estimating the errors of simple models fitted directly to the interferometric visibilities in the subsequent paper.

\item \textit{Small data sets.}

A problem may occur in snapshot observations when the residuals distribution is highly skewed. Then the observed residuals could poorly represent their population distribution. One can use the parametric bootstrap and sample the residuals from the Gaussian noise with the variance that has been estimated from the observed data.

\item \textit{Outlier visibility measurements.}

It is well known that a short-time amplitude error in the $uv$-domain results in ``waves'' of erroneous flux in the image-domain \citep{ekers}. Using such outlier visibility in resampling would result in replication of the outliers in the bootstrapped data samples. Most of these outlier visibility measurements are flagged by attributing zero or negative weight to them. But when they have a positive weight one should take care not to consider outliers in the resampling or stick to a parametric bootstrap with robustly estimated noise.

\end{itemize}

\subsection{Advantages}

One of the main advantages of using bootstrap is its wide applicability in assessing the uncertainties of VLBI results. The algorithm of errors estimation is the same for both the deconvolved image and the nonlinear combination of images or direct model of interferometric visibility. One does not need any other software except those used for image deconvolution/visibility fitting and the one used to generate the bootstrapped data sets.
In case of image-based estimates (VLBI maps and their non-linear combinations) the bootstrap allows one to account for the correlated and non-uniform noise in the image plane in contrast to the traditional $rms$-based approach. 
We also highlight the ability of the method to account for the inhomogeneous sensitivity of VLBI arrays, especially in the case of Space VLBI (e.g. \texttt{RadioAstron}, \citealt{radioastron}).
As already noted bootstrap can be extended beyond our implementation to account for self-calibration errors and other instrumental effects if these effects can be estimated.
Finally bootstrap can be used for hypothesis testing about any image or image combination structure, e.g. RM gradient.

\section{Conclusions}
\label{sec:conclusions}

%

Bootstrap is a well known and widely used method for assessing uncertainties in cases when the errors cannot be estimated analytically. However, up to now bootstrap has been completely ignored by the VLBI community, despite the fact that many widely-used algorithms of VLBI data processing lack uncertainty output. 
In this paper we demonstrated that bootstrap is a natural and efficient tool to estimate the uncertainties of the image-based statistics obtained using the inherently complex and non-linear VLBI data processing algorithms. It also naturally accounts for inhomogeneous sensitivity arrays (e.g. Space VLBI).

We conducted simulations with artificially generated data sets and found out that bootstrap provides error estimates for the images obtained using \texttt{CLEAN} and their combinations with better coverage than that obtained using traditional $rms$-based uncertainties.
Bootstrap allows one to formulate the criterion of a statistically significant Rotation Measure gradient in a image slice that is free of irrelevant points such as the width of the gradient or RM errors at some locations of the slice.

Using MOJAVE multi-frequency VLBA data we confirm that there are statistically significant (at level $\alpha$=0.05) transverse RM gradients in jets of 0923$+$392, 0945$+$408 and 2230$+$114 (Appendix~\ref{sec:real}). For 1641$+$399 the RM gradient is found to be statistically non-significant.

\section*{Acknowledgements}

The author would like to thank the anonymous referees for their thoughtful comments, which resulted in numerous improvements in this paper. 
The author also thanks Evgenya Kravchenko, Mikhail Lisakov and Alexander Kutkin for useful discussions, Evgenya Kravchenko,Yury Yu. Kovalev and, especially, Dimitry Litvinov for reading the manuscript and constructive suggestions.
This work is supported by Russian Science Foundation grant 16-12-10481.
This research has made use of data from the MOJAVE database that is
maintained by the MOJAVE team \cite{mojave6kinematicsanalysis}.

This research has made use of NASA's Astrophysics Data System.

This research made use of Astropy, a community-developed core Python package for Astronomy \citep{2013A&A...558A..33A}, \textit{Numpy} \citep{numpy} and \textit{Scipy} \citep{scipy}. \textit{Matplotlib} Python package \citep{Hunter:2007} was used for generating all plots in this paper.




\bibliographystyle{mnras}
\bibliography{paper} 




\appendix

\section{Re-analysis of the suspicious RM gradients}
\label{sec:real}

\begin{figure*}
   \centering
   \includegraphics[width=0.48\textwidth]{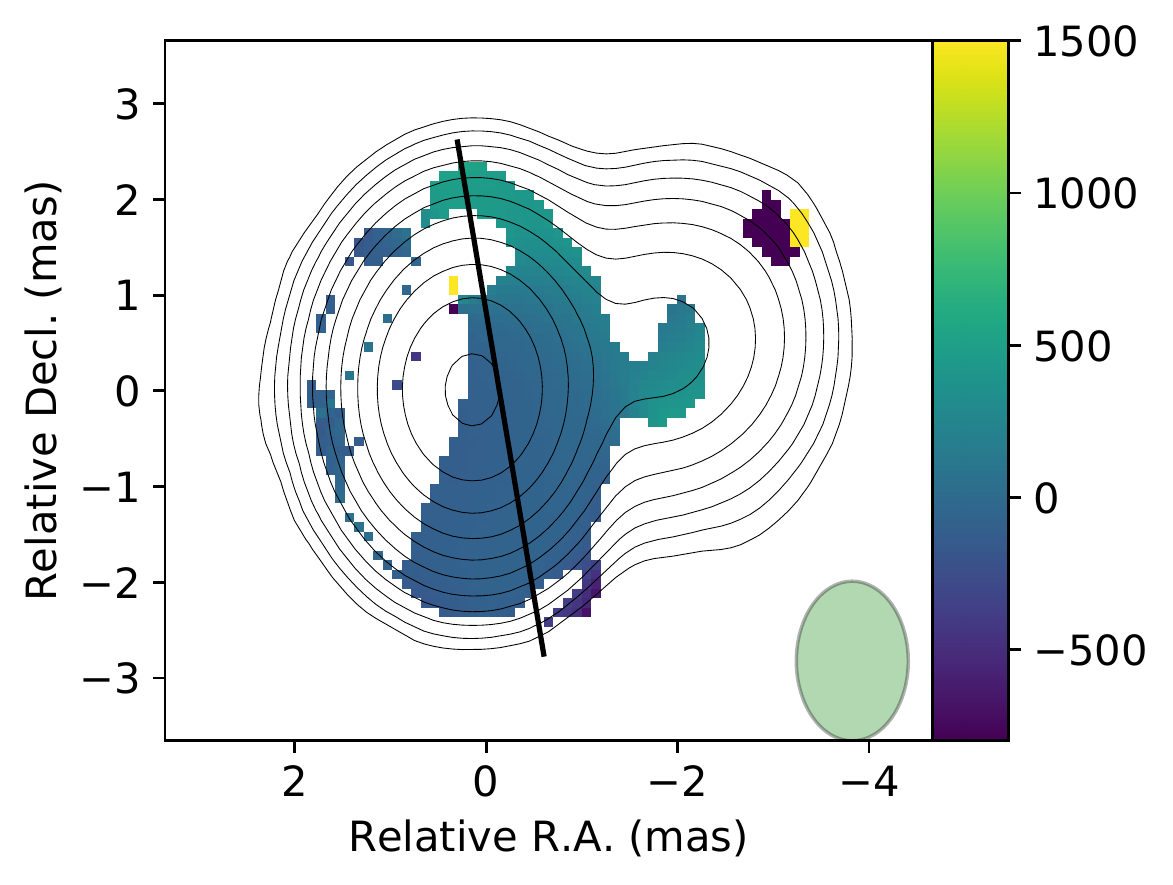}
   \includegraphics[width=0.48\textwidth]{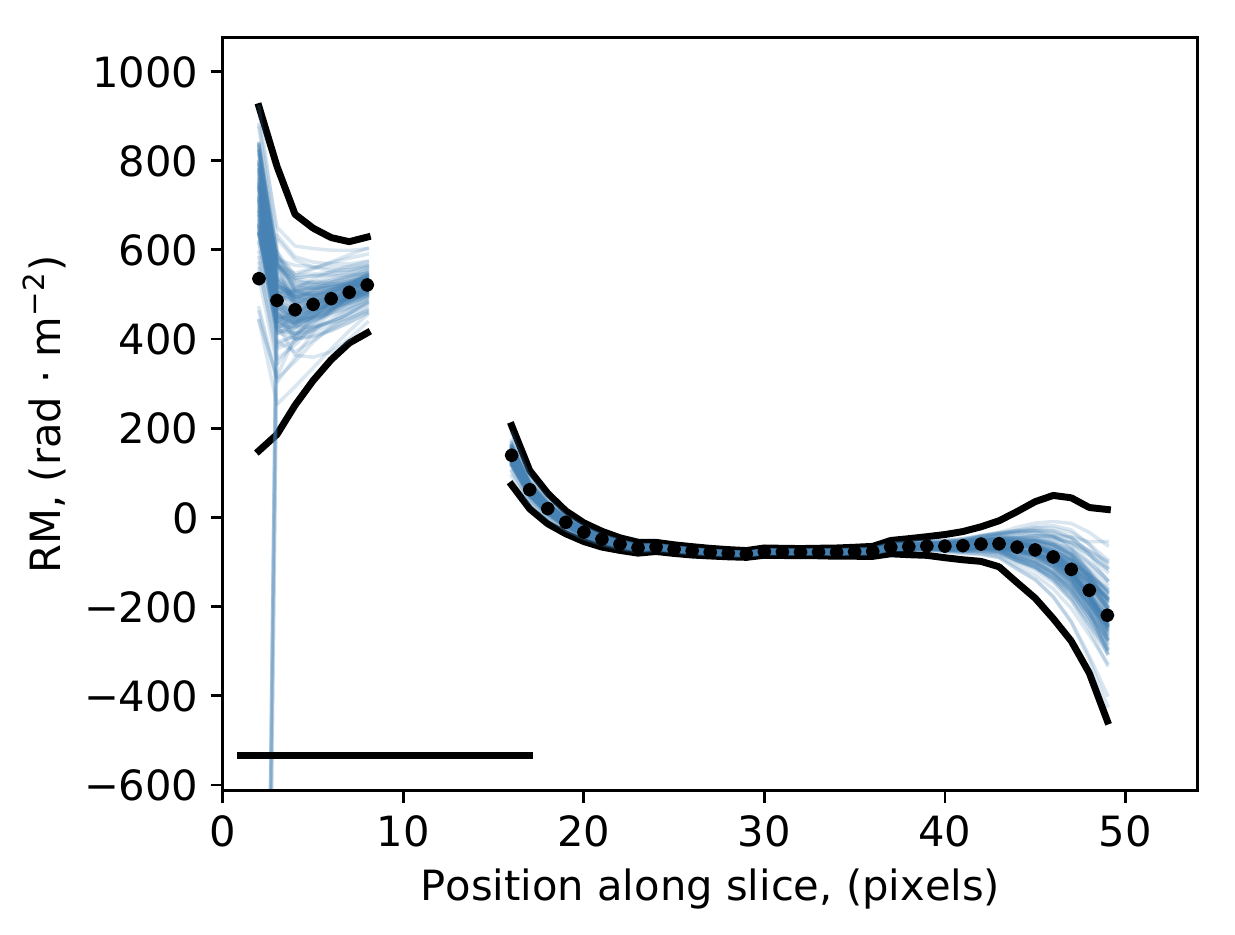}
      \caption{The rotation measure map (left) and its slice (right) with a significant transverse RM gradient for 0923$+$392. The green ellipse in the image and the straight black line in slice show the synthesized beam. The colour scheme is given in $\rm rad \cdot m^{-2}$. The black dots show the observed RM values, the blue thin lines -- bootstrap realizations and the black thick line -- SCB. Note that 5 bootstrap realizations have an exceptionally low RM values in the border point of slice due to low SNR.}
         \label{0923_grad}
   \end{figure*}

\begin{figure*}
   \centering
   \includegraphics[width=0.32\textwidth]{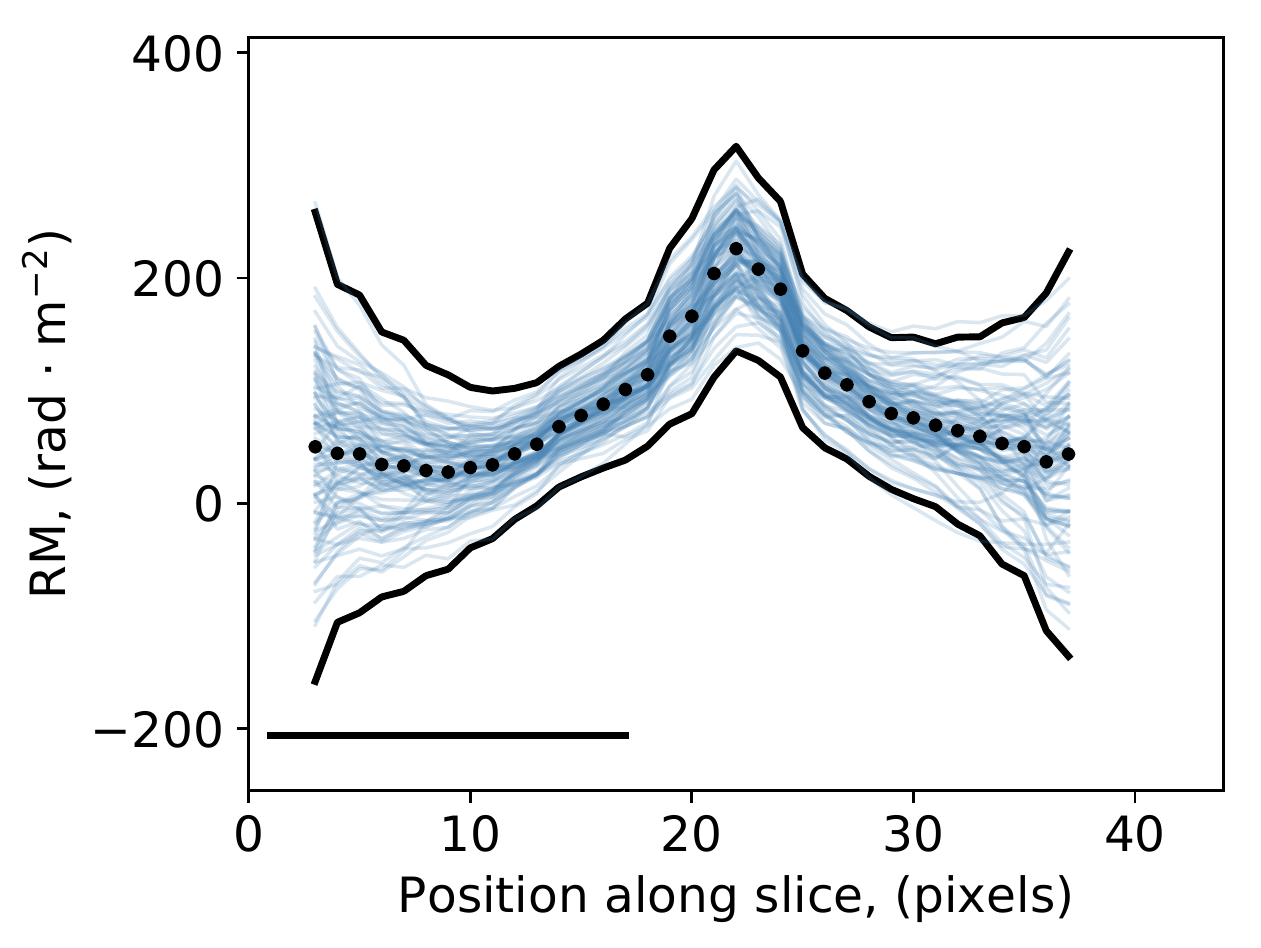}
   \includegraphics[width=0.32\textwidth]{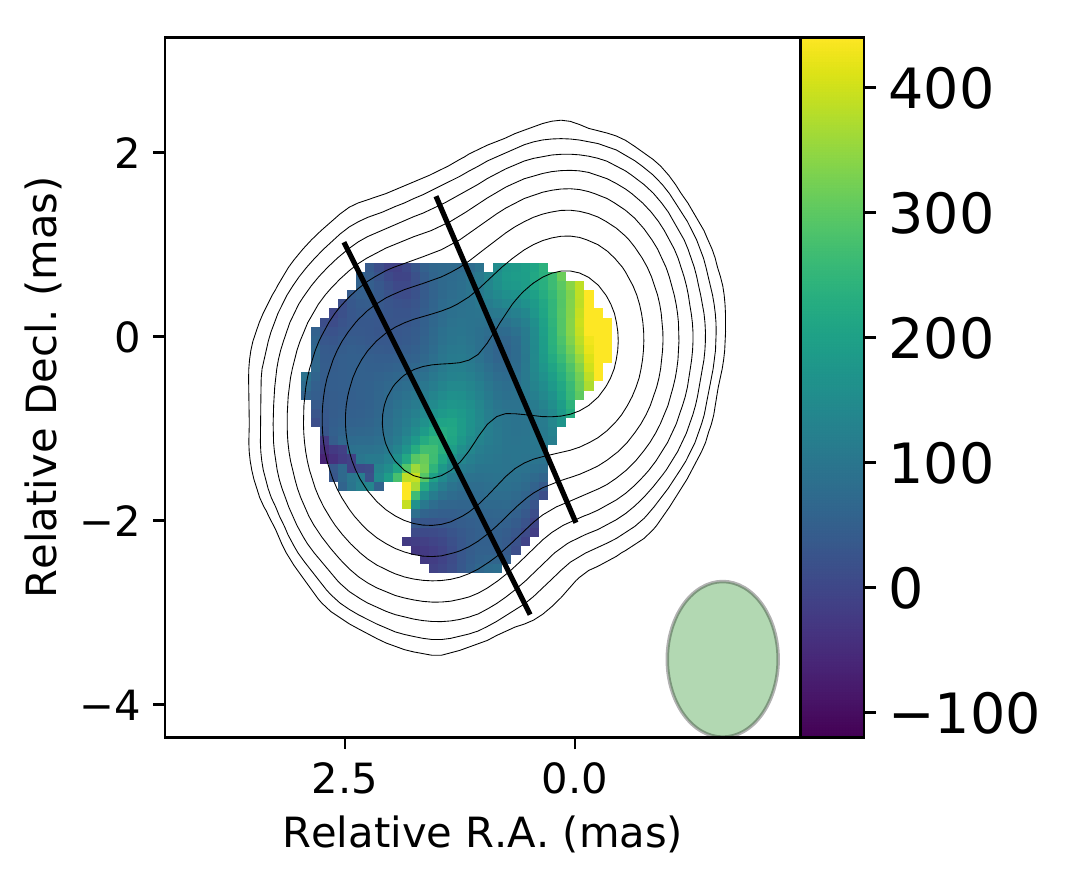}
   \includegraphics[width=0.32\textwidth]{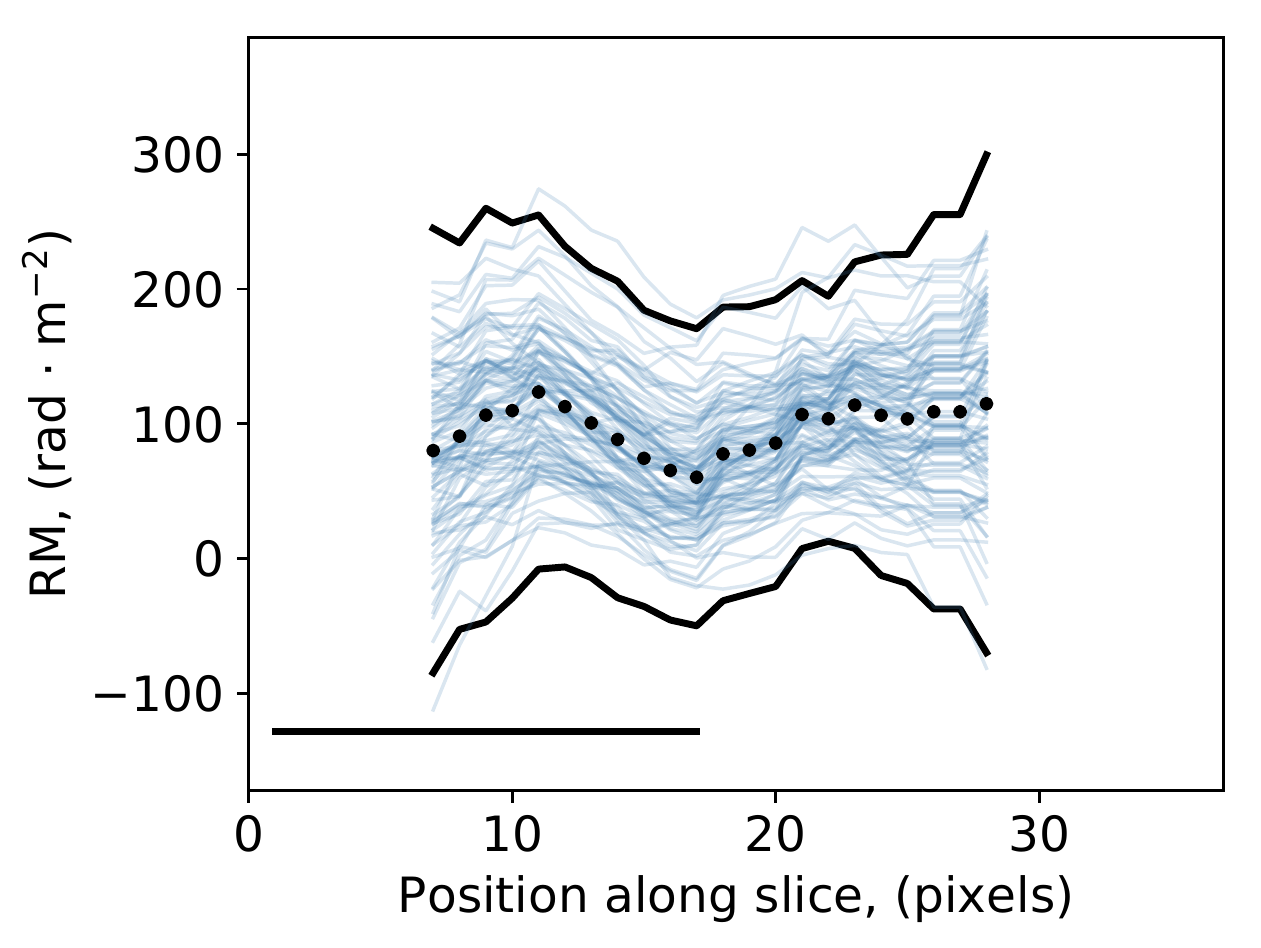}
      \caption{The rotation measure map (middle) and slices with significant (left) and non-significant (right) transverse RM gradients for 0945$+$408.}
         \label{0945_grad}
   \end{figure*}

\begin{figure*}
   \centering
   \includegraphics[width=0.48\textwidth]{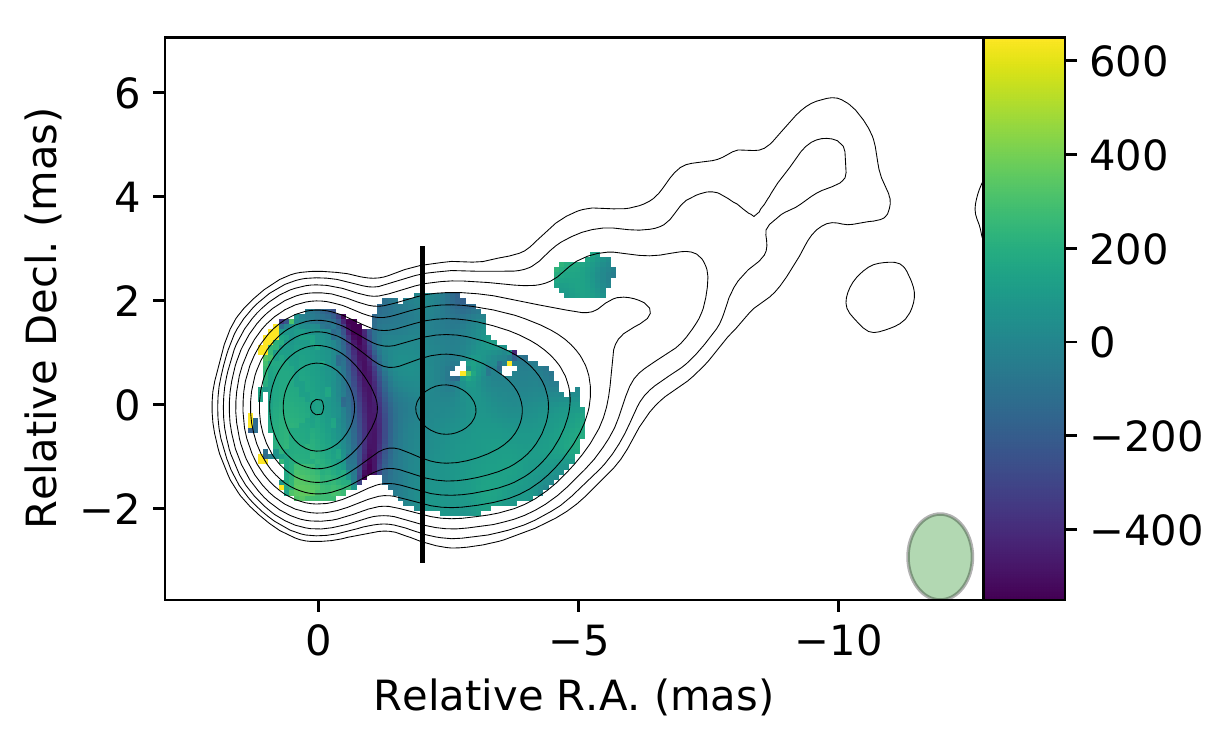}
   \includegraphics[width=0.48\textwidth]{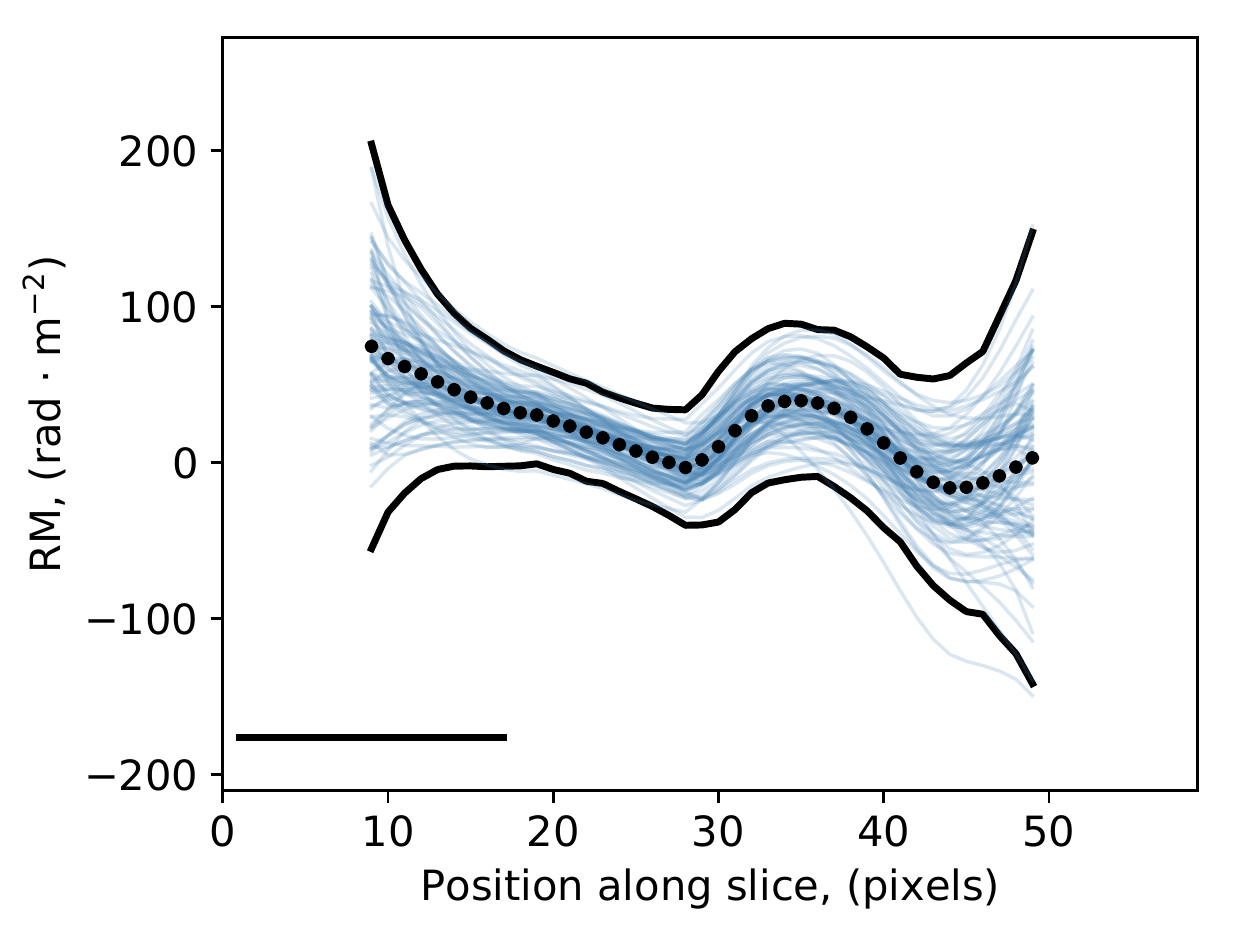}
      \caption{The rotation measure map and slice with non-significant RM gradient for 1641$+$399.}
         \label{1641_grad}
   \end{figure*}

We have applied the proposed criterion (see Section~\ref{sec:rmgrad}) to estimate the statistical significance of RM gradients for several sources from \cite{mojave8rm}. These sources were considered as probable candidates for significant gradients, but did not meet the strict  requirements stated by \cite{mojave8rm}. We found that there are statistically significant (at level $\alpha$=0.05, Section~\ref{sec:coverage}) transverse RM gradients in jets of 0923$+$392 (Fig.~\ref{0923_grad}), 0945$+$408 (Fig.~\ref{0945_grad}) and 2230$+$114 (Fig.~\ref{fig:rotm_slice}). For 1641$+$399 (Fig.~\ref{1641_grad}) the RM gradient is found to be statistically non-significant.

%


\bsp	
\label{lastpage}
\end{document}